\documentclass[journal]{IEEEtran}
\usepackage{ifpdf}
\usepackage{multirow}
\usepackage{multicol}
\usepackage{graphicx}
\usepackage{subfigure}
\usepackage{booktabs} 
\usepackage{amsmath}
\usepackage{amssymb}
\usepackage{float} 
\usepackage{microtype}
\usepackage{subfigure}
\usepackage{bm}
\DeclareMathOperator*{\argmax}{argmax}
\ifCLASSINFOpdf
\else
\fi
\begin{document}
\title{Specialized Decision Surface and Disentangled Feature for Weakly-Supervised Polyphonic Sound Event Detection
}

\author{Liwei~Lin$^{1,2}$,
        Xiangdong~Wang$^{1}$,
        Hong~Liu$^1$,
        and~Yueliang~Qian$^1$
        % <-this % stops a space
        \thanks{$^1$Bejing Key Laboratory of Mobile Computing and Pervasive Device, Institute of Computing Technology, Chinese Academy of Sciences, Beijing, China}
        \thanks{
        $^2$University of Chinese Academy of Sciences, Beijing, China}
        
}

\maketitle


%SED can be regarded as a supervised learning task when strong annotations (timestamps) are available during learning. However, due to the high cost of manual strong labeling data, it becomes crucial to introduce weakly supervised learning to SED, in which only weak annotations (clip-level annotations without timestamps) are available during learning.
\begin{abstract}
%Sound event detection (SED) consists in recognizing the presence of sound events in the segment of audio and detecting their onset as well as offset. In this paper, we focus on two common problems on SED: how to carry out efficient weakly-supervised learning and how to learn better from the unbalanced dataset in which multiple sound events often occur in co-occurrence.
In this paper, a special decision surface for the weakly-supervised sound event detection (SED) and a disentangled feature (DF) for the multi-label problem in polyphonic SED are proposed. We approach SED as a multiple instance learning (MIL) problem and utilize a neural network framework with a pooling module to solve it. General MIL approaches include two kinds: the instance-level approaches and embedding-level approaches. 
We present a method of generating instance-level probabilities for the embedding level approaches which tend to perform better than the instance-level approaches in terms of bag-level classification but can not provide instance-level probabilities in current approaches. Moreover, we further propose a specialized decision surface (SDS) for the embedding-level attention pooling. We analyze and explained why an embedding-level attention module with SDS is better than other typical pooling modules from the perspective of the high-level feature space.  As for the problem of the unbalanced dataset and the co-occurrence of multiple categories in the polyphonic event detection task,  we propose a DF to reduce interference among categories, which optimizes the high-level feature space by disentangling it based on class-wise identifiable information and obtaining multiple different subspaces. Experiments on the dataset of DCASE 2018 Task 4 show that the proposed SDS and DF significantly improve the detection performance of the embedding-level MIL approach with an attention pooling module and outperform the first place system in the challenge by $\mathbf{6.6}$ percentage points.
%We noted that compared with the original audio features, the number of frames of high-level features tends to be reduced by a series of pooling operations, which allows less trainable parameters and increases the receptive field of high-level features but also leads to coarser information contained in each frame-level feature.
%on the premise of no temporal scales compressed 
\end{abstract}

\begin{IEEEkeywords}
Sound event detection, machine learning, weakly-supervised learning, attention pooling.
\end{IEEEkeywords}

% \begin{center} \bfseries EDICS class: 3-BBND \end{center}
% \fi
\IEEEpeerreviewmaketitle

\section{Introduction}
\label{Introduction}
\IEEEPARstart{S}{ound} event detection (SED) is the task to detect and recognize individual sound sources in realistic soundscapes. It is required to recognize not only the presence of each event category in a sound source but also the start and end boundaries of each existing event. Since sounds carry a large amount of information about our everyday environment, SED supports many applications in everyday life, such as noise monitoring for smart cities \cite{bello2018sonyc}, bioacoustic species and migration monitoring \cite{stowell2015acoustic,salamon2016towards}, surveillance \cite{crocco2016audio}, healthcare \cite{goetze2012acoustic}, and large-scale multimedia indexing \cite{hershey2017cnn}. 

Polyphonic SED is often approached as a multi-label classification problem. Ideally, each event category to detect is modeled equivalently, regardless of the effect of co-occurrence of multiple events on each other. Moreover, a large amount of clean and balanced data with strong annotations (annotations with detailed timestamps for all event occurrences) is required for the model to train. However, in real life, the problem can be more complex. For example, noises out of the domain occurring in the audio recording increase the difficulty of the detection, unbalanced dataset and the co-occurrence of different events interfere with the detection of each other. Besides, it costs a lot to obtain large-scale strongly-annotated training data. Therefore, in this paper, noticing the difficulty to obtain large-scale strongly-annotated data and the problem caused by the co-occurrence of different events in the unbalanced dataset, we mainly focus on weakly-supervised SED and try to mitigate the impact of multi-event co-occurrence on the weakly-supervised SED system.

Weakly-supervised SED utilizes weak annotations, which only indicate the presence of event categories, rather than strong annotations during training. In this paper, we approach weakly-supervised SED as a multiple instance learning (MIL) problem and utilize a neural network framework with a pooling module to solve it. The MIL approaches provide strategies to decide the bag-level probability of a bag depending on all the instances in the bag. The pooling modules, such as global max pooling (GMP) \cite{oquab2015object}, global average pooling (GAP) \cite{zhou2016learning}, global weighted rank pooling (GWRP) \cite{kolesnikov2016seed,kong2018joint}, noisy-or pooling \cite{wang2018comparing} and attention pooling (ATP) \cite{ilse2018attention}, provide more detailed methods to integrate the instance-level probabilities into a bag-level probability or the instance-level feature representations into a bag-level contextual representation. The MIL approaches are distinguished as the instance-level approach and the embedding-level approach \cite{ilse2018attention}. Since Ilse et al. \cite{ilse2018attention} point out that the embedding-level approach is superior to the instance-level approach in terms of bag-level classification and most of the current work on weakly-supervised SED mainly focus on the instance-level approaches, we illustrate and demonstrate the effectiveness of the embedding-level approach on weakly-supervised SED. Since the embedding-level approach does not provide the instance-level probabilities, we propose a shared decision surface for it to generate the instance-level probabilities. We argue that the classifier learning from the bag-level contextual representation forms a shared decision surface of both the bag-level contextual representation and the instance-level high-level feature representations, for which the instance-level probabilities could be obtained by passing the instance-level high-level feature representations through the same classifier directly. Then we are able to compare the embedding-level performances of different MIL approaches cooperating with different pooling modules.

Furthermore, we propose a specialized decision surface to make better instance-level predictions than the shared decision surface for the embedding-level ATP.
Compared to other pooling modules without trainable parameters such as GAP and GMP, ATP is more flexible and potential to learn detailed instance-level information. We argue that the trainable parameters in the embedding-level ATP imply a better instance-level decision surface, namely the proposed specialized decision surface, for the embedding-level ATP to give the instance-level probabilities. We demonstrate that the embedding-level approach with the shared decision surface tends to perform better than the instance-level approach and the proposed specialized decision surface is more conducive to instance-level classification than the shared decision surface on the experimental dataset.

Besides, to mitigate the impact of multi-event co-occurrence of the unbalanced dataset on the proposed weakly-supervised polyphonic SED system, we also propose a disentangled feature, which re-models the high-level feature space so that the feature subspace of a certain category differs from other categories without pre-training. Since the disentangled feature re-models both the instance-level and the bag-level high-level feature space, it aims at the multi-label problem of both the weak labels and the overlapping sound events of the strong labels for polyphonic SED. The volume of these disentangled feature subspaces depends on the number of available clips containing strong class-wise identifiable information with little interference from other categories. In virtue of the introduction of more class-wise prior information as well as network redundancy weight reduction, the disentangled feature is able to improve the performance of the polyphonic SED system.

Our experiments show that the embedding-level ATP with specialized decision surface and disentangled feature outperforms other pooling modules as well as simple embedding-level ATP. Detailed analysis of the high-level feature space in the experiment also supports our hypothesis. 

The rest of this paper is organized as follows. We introduce related work about weakly-supervised polyphonic SED in Section~\ref{Related work}, describe in detail the MIL framework with different pooling modules in Section~\ref{MIL}, introduce the proposed methods in Section~\ref{Methods}, describe the dataset and configuration of our experiments in Section~\ref{Experiments}, analyze the results of experiments in Section~\ref{discussion} and draw conclusions in Section~\ref{Conclusions}.

%Furthermore, we propose a disentangled feature (DF) to improve multi-category learning. In polyphonic SED, there are multiple events occurring in co-occurrence at clip-level and highly overlapping with each other at frame-level in some clips. Due to the unbalanced data set and the fact that multiple categories occurs in co-occurrence at clip-level, it is difficult for cATP to learn multiple separate feature subspaces for those event categories that are highly overlapped with other event categories, especially for those with relatively few occurrences. Therefore, by taking into account the category overlap information, we propose a disentangled feature which re-models the high-level feature space so that the feature subspace of a certain category differs from other categories without pre-training. The scales of these disentangled feature subspaces depend on the number of available clips containing strong class-wise identifiable information with less interference from other categories. In virtue of the introduction of more class-wise prior information as well as network redundancy weight reduction, the disentangled feature can be regarded as a regularization method to help improve the performance of cATP-MIL frameworks.

%\IEEEpubidadjcol

\section{Related work}
\label{Related work}
\subsection{Multiple instance learning}
Weakly supervised learning is often approached as an MIL problem \cite{maron1998framework,dietterich1997solving}. It is especially common in medical image \cite{quellec2017multiple,xu2014deep} and semantic segmentation \cite{papandreou2015weakly,wu2014milcut}. The excellent performance of neural networks in various fields promotes the combination of the MIL framework and neural networks for weakly supervised learning \cite{pathak2014fully,zhou2002neural,wu2015deep,kraus2016classifying}. According to MIL, a bag of several instances has only bag-level annotations, in other words, does not have instance-level annotations. If there is at least one positive instance in a bag, the bag is annotated as a positive bag. Otherwise, the bag is annotated as a negative bag. The combination of the MIL framework and neural networks focuses on how to integrate several instance-level outputs of neural networks into a bag-level output so as to enable the model to calculate loss with only clip-level annotations and carry out end-to-end learning. Since neural networks have been widely used as a general high-level feature extractor in various tasks, the MIL framework with neural networks typically comprises a neural network feature extractor which generates the high-level feature representation sequence and a pooling module such as GMP, GAP, GWRP and ATP, which integrates instance-level outputs into a bag-level output.

\begin{figure}[t]
  \centering
  \subfigure[Instance-level approach]{\label{figp1a}\includegraphics[width=0.4\columnwidth]{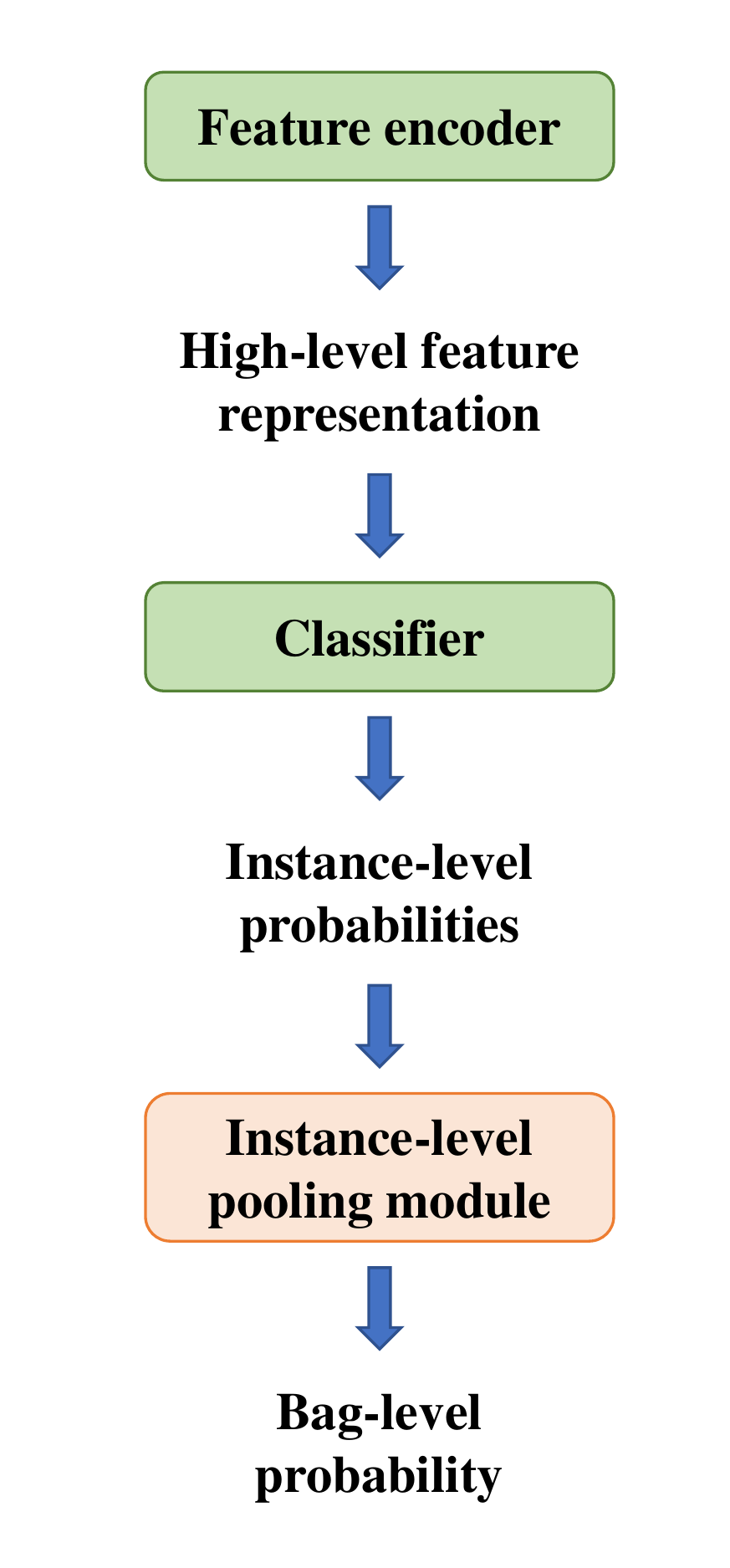}}
  \subfigure[Embedding-level approach]{\label{figp1b}\includegraphics[width=0.4\columnwidth]{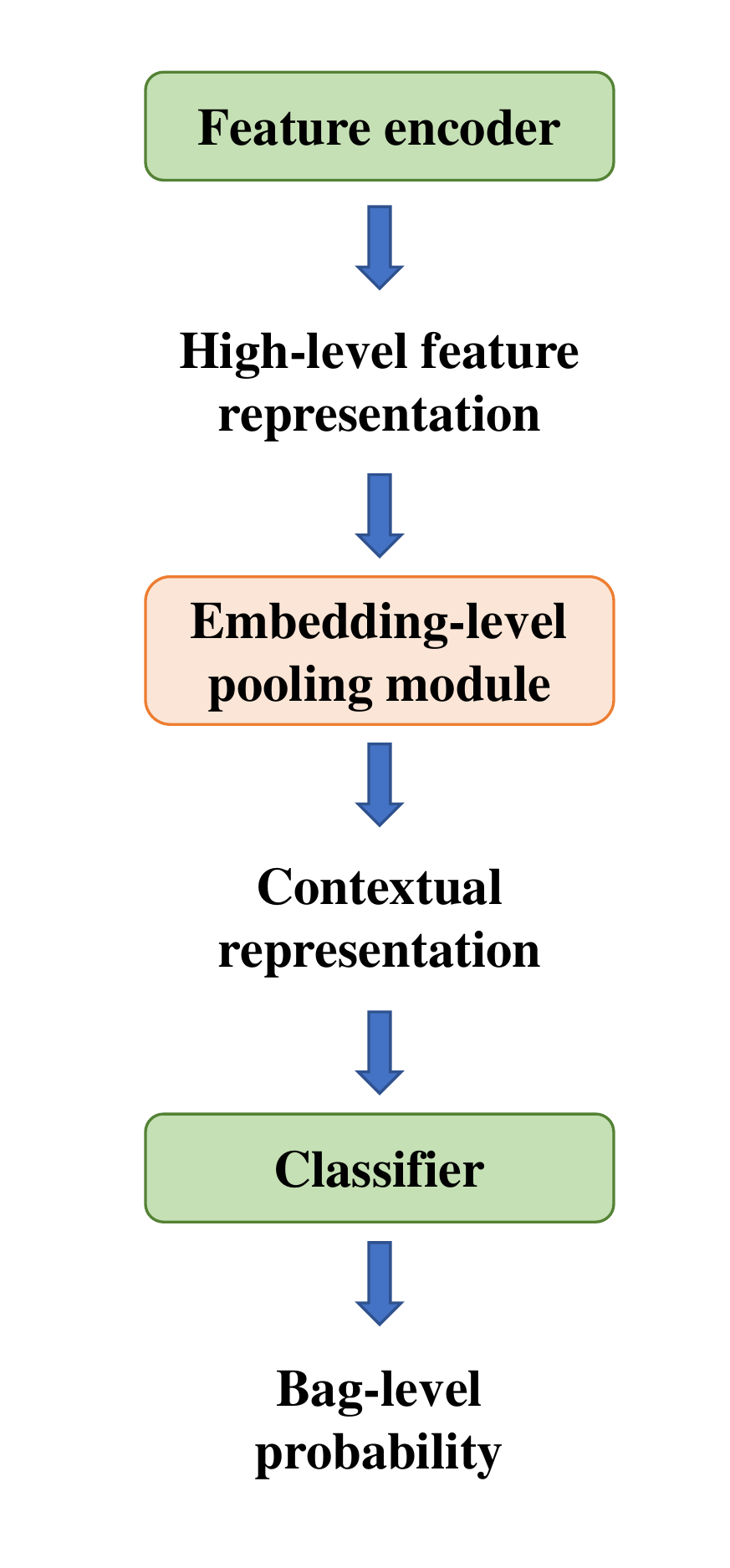}}
  \caption{The comparison of the instance-level approach and the embedding-level approach.}
  \label{figp1}
  \vskip -0.2in
\end{figure}

As mentioned in \cite{ilse2018attention} and shown in Figure~\ref{figp1}, the MIL approaches with neural network are distinguished as instance-level approaches and embedding-level approaches according to whether the pooling module involved in the MIL framework integrates instance-level probabilities into a bag-level probability or integrates instance-level high-level feature representations into a bag-level high-level feature representation. Ilse et al. \cite{ilse2018attention} also point out that the embedding-level approach is superior to the instance-level approach in terms of bag-level classification and proposes an attention-based embedding-level approach, which performs the best compared to other MIL approaches.

%In this paper, we provide a method for all the embedding-level approaches to generate instance-level probabilities. The proposed method considers that the instance-level high-level feature representations can share the same decision surface with the bag-level contextual representation, and we utilize a shared classifier in implementation. Furthermore, we explore the nature of the ability of the attention-based embedding-level approach to indicate key instances and thereby propose a method to generate more accurate instance-level probabilities for the attention-based embedding-level approach. Inspired by how the shared decision surface forms during training, the proposed method focuses on the forming of another latent specialized decision surface different from the shared decision surface during learning attention weight, based on which we re-design a classifier determined by this specialized decision surface and take probabilities output by this classifier as instance-level probabilities.

\subsection{Weakly-supervised sound event detection}
As for weakly-supervised SED, if we treat each frame in an audio clip as an instance, then the audio clip can be regarded as a bag with clip-level annotations (without frame-level annotations of frames). If a sound event occurs in any frame of the audio clip, the clip is considered as a positive clip of the event. Otherwise, the audio clip is treated as a negative audio clip. 

Since a pooling module described in the previous section is essential to an MIL framework with neural network, there are lots of previous works about MIL with different pooling modules for weakly-supervised SED: a fully convolutional network with a GMP module \cite{su2017weakly}, a joint detection-classification (JDC) model with a ATP module \cite{kong2017joint}, a convolutional neural network (CNN) based model with a GAP module \cite{kumar2018knowledge}, a gated convolutional recurrent neural network (CRNN) with a global softmax pooling (GSP) module \cite{xu2018large} and a joint separation-classification (JSC) model with a GWRP module \cite{kong2018joint}. Especially, McFee et al. \cite{mcfee2018adaptive} explore the effects of different pooling modules such as GAP, GMP and GSP and propose an adaptive pooling module. Wang et al. \cite{wang2019comparison} offer a comparison of several pooling modules including GAP, GMP, GSP and ATP.

However, all these work described above just considers the instance-level approaches. To explore the effects of different pooling modules cooperating with both of instance-level and embedding-level approaches on SED, we carry out a series experiments and find that the embedding-level approach tends to perform better.

%We describe in detail these effects in Section~\ref{discussion}.
%Therefore, in this paper, we describe in detail 4 instance-level pooling approaches and 4 embedding-level pooling approaches and demonstrate that the embedding-level attention pooling module combined with the proposed specialized decision surface performs best on the experimental dataset.
%for more accurate boundary detection for SED.

\subsection{Polyphonic sound event detection}

Since multiple event categories tend to occur in co-occurrence in an audio clip, a SED system is also termed as a polyphonic SED system \cite{cakir2015polyphonic,parascandolo2016recurrent,cakir2017convolutional}. Recently, neural networks such as recurrent neural network (RNN) \cite{hayashi2017duration} and CRNN \cite{cakir2017convolutional} show a significant effect on polyphonic SED. Commonly, these methods model each event category equally. When designing models, they make all the categories share the same feature encoder. However, in realistic sound environment, multiple events overlapping in the unbalanced dataset, such as ``Dishes" and ``Frying", would interference with the recognition of each other, especially when the numbers of clips of some event categories are relatively small and thereby less  identifiable information about these event categories can be available. During training, the feature encoder tends to fit better for some event categories with more identifiable information than those with little identifiable information. To tackle this problem, Imoto et al. \cite{imoto2019sound} proposes a neural-network-based SED with graph Laplacian regularization based on the co-occurrence of sound events. We also focus on this problem and try to take advantage of more prior information about the data distribution to optimize the high-level feature space of the feature encoder, thereby making more accurate classification and detection for overlapping events.

\section{MIL for weakly-supervised polyphonic SED}
\label{MIL}
In this section, we describe in detail the MIL framework for weakly-supervised polyphonic SED. 8 common pooling modules including 4 instance-level pooling modules and 4 embedding-level pooling modules are introduced.

\subsection{The MIL framework}
\label{Multiple instance learning for polyphonic SED}
For weakly-supervised polyphonic SED, since multiple different events might appear in co-occurrence in the same audio clip, we consider each event category separately when approaching SED as an MIL problem. Assuming that there are $C$ event categories to detect, then for event category $c$, we treat an audio clip as a positive audio clip if the audio clip contains event category $c$. Otherwise, the audio clip is treated as a negative audio clip. Let $\mathbf{x}=\left\{ \mathbf{x}_1,\mathbf{x}_2,\ldots,\mathbf{x}_T\right\}$ be the high-level feature representations of the audio clip generated by the feature encoder and $\mathbf{y}=\left\{y_1,y_2,\ldots,y_C\right\}$ ($y_c\in\left\{0,1\right\}$) be the groundtruths, where $C$ is the number of categories.

For the instance-level approach, for each event category $c$, the high-level feature representations are passed into the classifier to generate frame-level probabilities $\hat{p}\left(y_c\mid \mathbf{x}_1\right),
\hat{p}\left(y_c\mid \mathbf{x}_2\right),
\ldots,\hat{p}\left(y_c\mid \mathbf{x}_t\right)\ldots,
\hat{p}\left(y_c\mid \mathbf{x}_T\right)$. Then the instance-level pooling module aggregates frame-level probabilities into a clip-level probability:

\begin{equation}
\hat{P}\left(y_c\mid\mathbf{x}\right)=
{\rm POOLING}\{
\hat{p}\left(y_c\mid \mathbf{x}_1\right),
\ldots,
\hat{p}\left(y_c\mid \mathbf{x}_T\right)
\}.
\end{equation}
When making predictions, assuming that $\alpha$ is a threshold for clip-level prediction and $\gamma$ is a threshold for frame-level prediction. Then the clip-level prediction is:

\begin{equation}
\phi_{c}\left(\mathbf{x}\right)=\left\{\begin{matrix}
1,&\hat{P}\left(1\mid\mathbf{x}\right)\geq \alpha \\ 
0,&{\rm otherwise}
\end{matrix}\right.
,
\label{eq2}
\end{equation}
where $\hat{P}\left(1\mid\mathbf{x}\right)$ denotes the probability that the clip is predicted to be positive.
The the frame-level prediction at time $t$ is:
\begin{equation}
\varphi_{c}\left(\mathbf{x},t\right)
=\left\{\begin{matrix}
1,&\hat{p}\left(1\mid \mathbf{x}_t\right)\cdot\phi_{c}\left(\mathbf{x}\right)\geq \gamma \\ 
0,&{\rm otherwise}
\end{matrix}\right.
,
\label{eq3}
\end{equation}
where $\hat{p}\left(1\mid \mathbf{x}_t\right)$ denotes the probability that the $t^{\mathrm{th}}$ frame is predicted to be positive. Without loss of generality, we set $\alpha=0.5$ and $\gamma=0.5$ in our experiments. 

For the embedding-level approach, the embedding-level pooling module directly aggregates all the high-level feature representations into a contextual representation $\mathbf{h}_c$:

\begin{equation}
\mathbf{h}_c=
{\rm POOLING}\left(\mathbf{x}_1,\mathbf{x}_2,\ldots,\mathbf{x}_T\right).
\end{equation}
Then the clip-level probability can be obtained by passing the contextual representation into the classifier:

\begin{equation}
\hat{P}\left(y_c\mid\mathbf{x}\right)=\hat{P}\left(y_c\mid \mathbf{h}_c\right)=G_c(\mathbf{h}_c),
\label{eq5}
\end{equation}
%\begin{figure}[t]
 % \centering
 % \includegraphics[width=\columnwidth]{tu1.pdf}
 % \caption{ The comparison of embedding-level pooling module and instance-level pooling module in MIL framework.}
 % \label{fig1}
%\end{figure}
where $G_c$ is the classifier of $\mathbf{h}_c$ to generate the clip-level probability.
Therefore, the clip-level prediction for the embedding-level approach can be obtained according to Equation~\ref{eq2} and \ref{eq5}.

\subsection{Instance-level pooling modules}
\label{Instance-level pooling modules}
We introduce $4$ typical pooling modules for the instance-level MIL, namely GMP, GAP, GSP, and ATP. These $4$ instance-level pooling modules are commonly used in weakly SED, such as GMP in \cite{su2017weakly}, GAP in \cite{kumar2018knowledge}, GSP in \cite{xu2018large,yan2019region} and ATP in \cite{kong2017joint}.

For GMP, the clip-level probability only depends on the maximum probability of all the frame-level probabilities of an audio clip:

\begin{equation}
\hat{P}\left(y_c\mid\mathbf{x}\right)=
\underset{t}{\max}\; \hat{p}\left(y_c\mid \mathbf{x}_t\right).
\end{equation}
\begin{figure*}[t]
  \centering
  \subfigure[GMP]{\label{fig2a}\includegraphics[width=0.38\columnwidth]{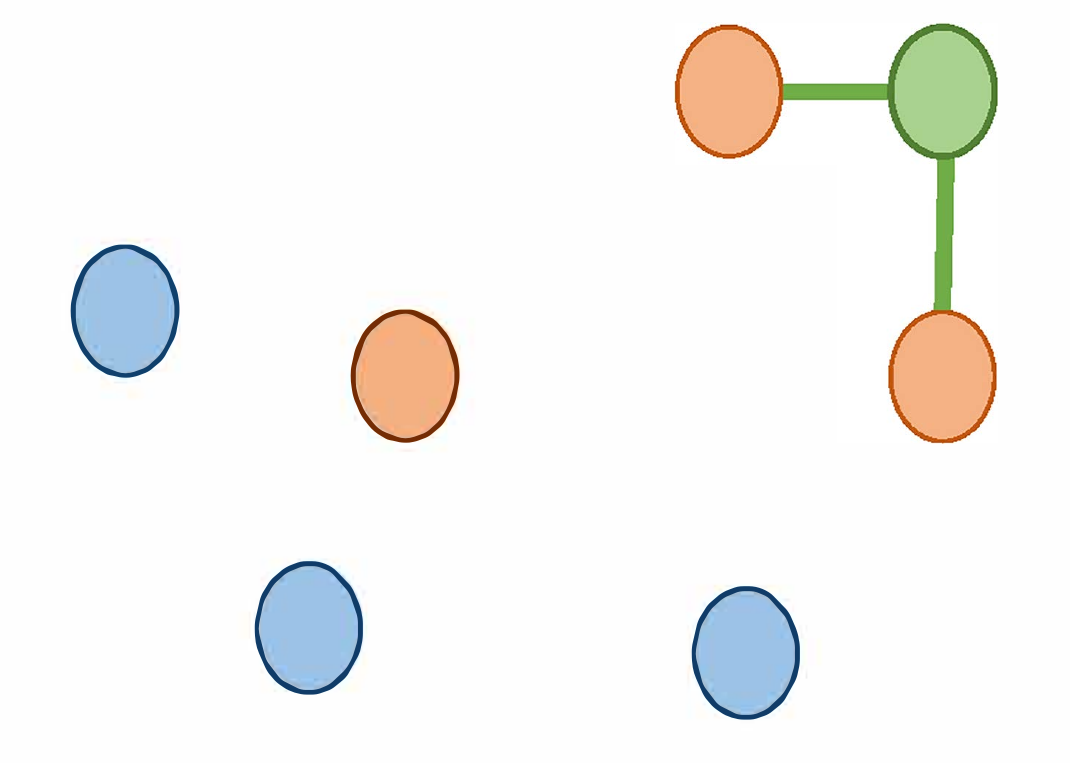}}
  \subfigure[GAP]{\label{fig2b}\includegraphics[width=0.38\columnwidth]{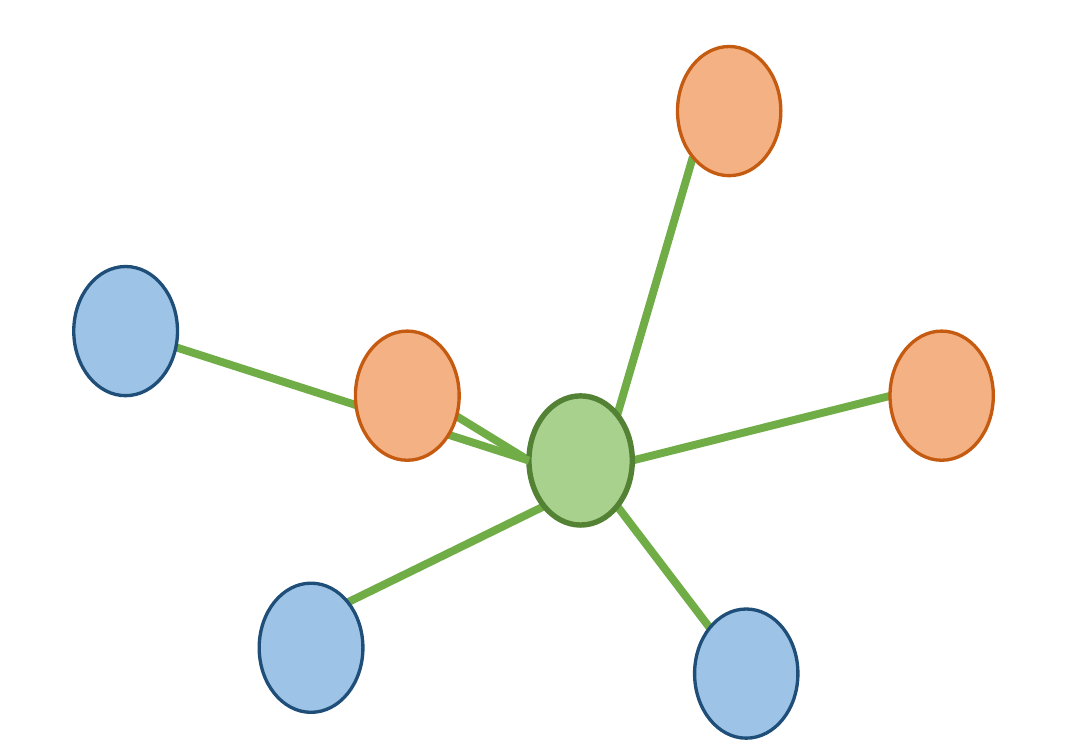}}
  \subfigure[GSP]{\label{fig2c}\includegraphics[width=0.38\columnwidth]{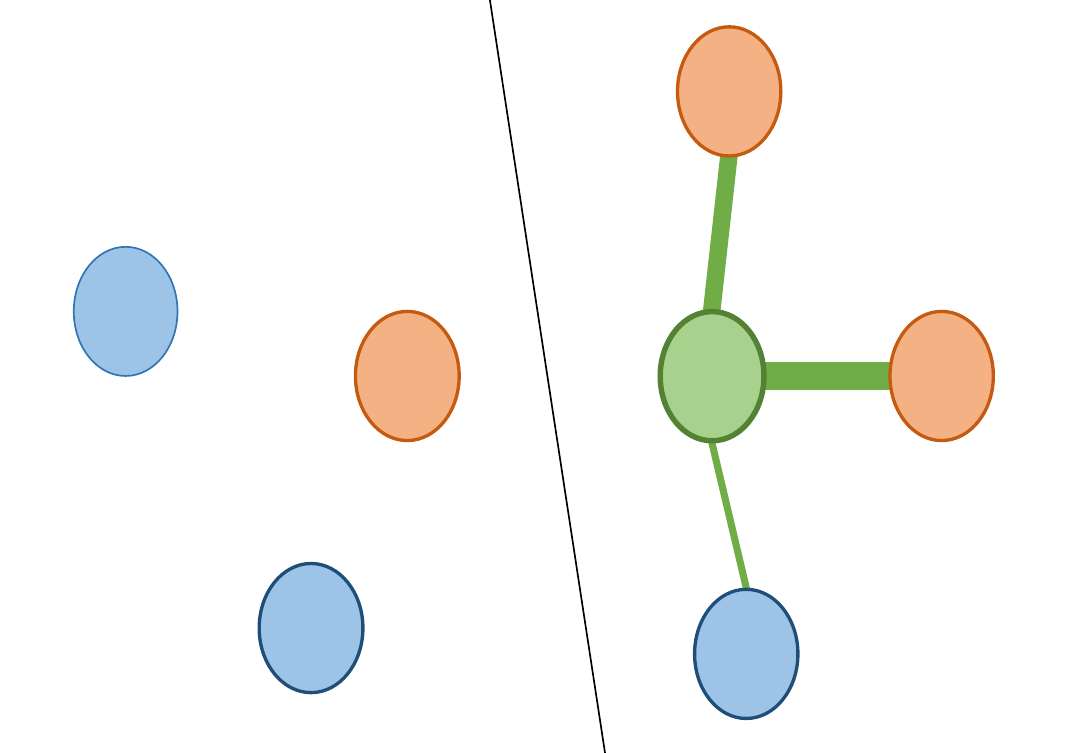}}
  \subfigure[ATP]{\label{fig2d}\includegraphics[width=0.78\columnwidth]{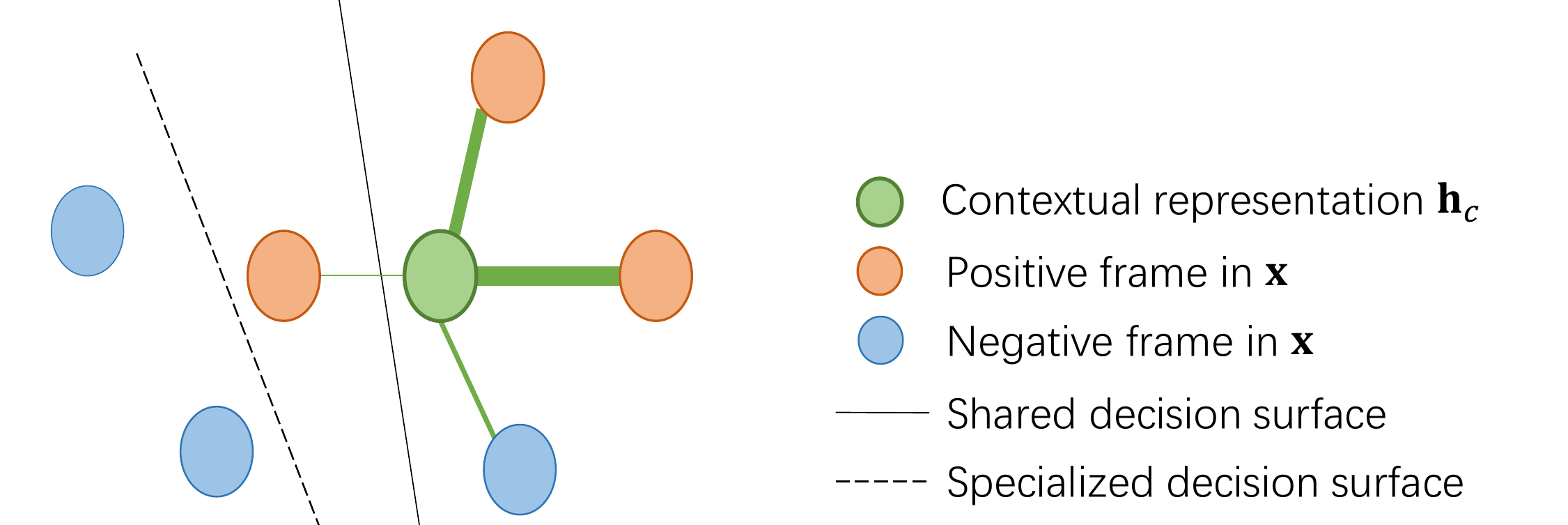}}
  \caption{A sketch of the relation of $\mathbf{h}_c$ and $\mathbf{x}$ of a positive audio clip in the 2-dimension feature space for event category $c$. }
  \label{fig2}
  \vskip -0.15in
\end{figure*}
For GAP, the clip-level probability relates to all the frame-level probabilities of an audio clip. More specifically, it takes the average value of all the frame-level probabilities as clip-level probabilities:

\begin{equation}
\hat{P}\left(y_c\mid\mathbf{x}\right)=
\frac{1}{T}\: \underset{t}{\sum}\:  
\hat{p}\left(y_c\mid \mathbf{x}_t\right).
\end{equation}

Obviously, since $\hat{P}\left(y_c\mid\mathbf{x}\right)$ only relates to the $s^{th}$ high-level feature representation $\mathbf{x}_s$ ($s=\underset{t}{\argmax}\; \hat{p}\left(y_c\mid \mathbf{x}_t\right)$), GMP only updates a limited number of weights of the neural network for each clip. On the other hand, although GAP considers all the high-level feature representations when updating the neural network, it focuses on each frame equally, ignoring the different degrees how much a frame contributes to the audio clip. Then a weighted pooling is proposed to fix this defect:

\begin{equation}
\hat{P}\left(y_c\mid\mathbf{x}\right)=\sum_t a_{ct}\cdot \hat{p}\left(y_c\mid \mathbf{x}_t\right),
\label{eq8}
\end{equation}
where $a_{ct}$ denotes the contribution of the $t^{th}$ frame to an audio clip for event category $c$.

GSP and ATP are two examples of such weighted pooling modules, where GSP connects the contribution of frames with frame-level probabilities:

\begin{equation}
 a_{ct}=\frac{\exp\left(
 \psi\left(\hat{p}\left(y_c\mid \mathbf{x}_t\right)\right)
 \right)}{\sum_k \exp\left(\psi\left(\hat{p}\left(y_c\mid \mathbf{x}_k\right)\right)\right)},
\label{eq9}
\end{equation}
where $\psi$ is a function to scale $\hat{p}\left(y_c\mid \mathbf{x}_t\right)$ appropriately.

Different from GSP, ATP offers an independent detector to generate the contribution of frames. This independent detector is learnable and in this paper, we give a common form of such an independent detector:

\begin{equation}
a_{ct}=\frac{\exp\left( \mathbf{w}_c^{\mathsf{T}} \mathbf{x}_t/d\right)}{\sum_k \exp\left(\mathbf{w}_c^{\mathsf{T} }\mathbf{x}_k/d\right)},
\label{eq10}
\end{equation}
where $\mathbf{w}_c$ are learnable parameters of the given independent detector and $d$ is a scaling factor to avoid too-large value of $\mathbf{w}_c^{\mathsf{T}}\mathbf{x}_t$. The value of $d$ is generally consistent with the dimensions of $\mathbf{x}_t$.

\subsection{Embedding-level pooling modules}
\label{Embedding-level pooling modules}
We describe how the $4$ pooling modules discussed above cooperate with the embedding-level MIL approach. In fact, though the embedding-level MIL approach is introduced and claimed to be superior to the instance-level MIL approach in \cite{ilse2018attention}, its application in SED is rare.

Different from the instance-level pooling modules, the embedding-level pooling modules work by integrating high-level feature representations $\mathbf{x}$ instead of frame-level probabilities $\hat{p}\left(y_c\mid \mathbf{x}_t\right)$ as described in Section~\ref{Multiple instance learning for polyphonic SED}. 

For GMP, assuming that the contextual representation $\mathbf{h}_c=\left\{h_{c1},h_{c2},\ldots,h_{cZ}\right\}$ is an $Z$-dimensional vector and $x_{tz}$ is the $z^{th}$ component of $\mathbf{x}_t$, then the $z^{th}$ component of the contextual representation $\mathbf{h}_c$ is
\begin{equation}
%h_e=\underset{t}{\max}\; {\mathbf{x_t}}_e
%\label{eq11}
h_{cz} = \max\; \{x_{1z}, x_{2z}, \ldots, x_{tz} \ldots, x_{Tz}\}.
\label{eq11}
\end{equation}

For GAP, the contextual representation $\mathbf{h}_c$ is:
\begin{equation}
\mathbf{h}_c=\frac{1}{T}\: \underset{t}{\sum}\: \mathbf{x}_t.
\end{equation}

For GSP and ATP, the contextual representation $\mathbf{h}_c$ is:

\begin{equation}
 \mathbf{h}_c=\sum_t a_{ct}\cdot \mathbf{x}_t.
 \label{eq13}
\end{equation}
Similar to instance-level pooling modules, $a_{ct}$, the contribution of $\mathbf{x}_t$ to an audio clip for event category $c$ is attained by Equation~\ref{eq9} for GSP and by Equation~\ref{eq10} for ATP.

\section{Methods}
\label{Methods}
In this section, we propose how to generate frame-level probabilities for the embedding-level approach and introduce the proposed specialized decision surface (SDS) and disentangled feature (DF). 

\subsection{Shared decision surface}
\label{Shared decision surface}

Since there are no frame-level probabilities generated during learning, we propose that the clip-level contextual representation and frame-level high-level feature representations can share the same classifier, despite the fact that the classifier is simply utilized for classification of the contextual representation during training. 

% as shown in Figure~\ref{fig0}
%which makes the formation of frame-level probabilities consistent with that of the instance-level approach.

%while the frame-level prediction for event $c$ at time $t$ can be obtained by Equation~\ref{eq3}.

We argue that not only the model learns the decision surface (the classifier) explicitly for $\mathbf{h}_c$ but also learn a latent decision surface for $\mathbf{x}$. Since this latent decision surface can not be obtained directly, we consider it to be close to the decision surface of $\mathbf{h}_c$.

As shown in Figure~\ref{fig2}, to simplify the analysis, we assume that the high-level feature space is a 2-dimension space and sketch the relation of $\mathbf{h}_c$ and $\mathbf{x}$ in this 2-dimension feature space for event category $c$. The green circles represent $\mathbf{h}_c$, the orange circles represent positive frames in $\mathbf{x}$ and the blue circles represent negative frames in $\mathbf{x}$. 
We connect $\mathbf{h}_c$ and $\mathbf{x}$ in the following way: draw a line between $\mathbf{h}_c$ and a frame in $\mathbf{x}$ if there is a connection between them and let the thickness of the line corresponding to the frame indicates the strength of the connection. 

We assume that the decision surface is fixed, and explore how the feature encoder tends to form a high-level feature space to fit the decision surface. The relative position of the high-level feature representation of each frame to the decision surface changes constantly with the formation of the feature space. During training, the green circle of a positive audio clip tends to move toward the positive side of the decision surface, which implies that those circles connected with the green circles move with the green circle together. On the contrary, those circles connected with the green circle in a negative audio clip tends to move toward the negative side of the decision surface. 

If circles (except green circles) with such a connection are considered to be positive and the strength of the connection is related to the confidence that it is considered positive, then the movement in a positive audio clip carries these (both true-positive and false-positive) circles toward the positive side of the decision surface. Since there is no true-positive circle in a negative audio clip, the movement in this circumstance actually carries false-positive circles toward the negative side of the decision surface. Attribute to these movements, positive (true-positive) frames and negative (false-positive) frames are able to gradually separate into two clusters and the decision surface of the contextual representations is close to the boundary of such two clusters, in other words, suitable for coarse frame-level classification, for which we regard it as a shared decision surface.

\begin{figure}[t]
 \centering
  \includegraphics[width=0.6\columnwidth]{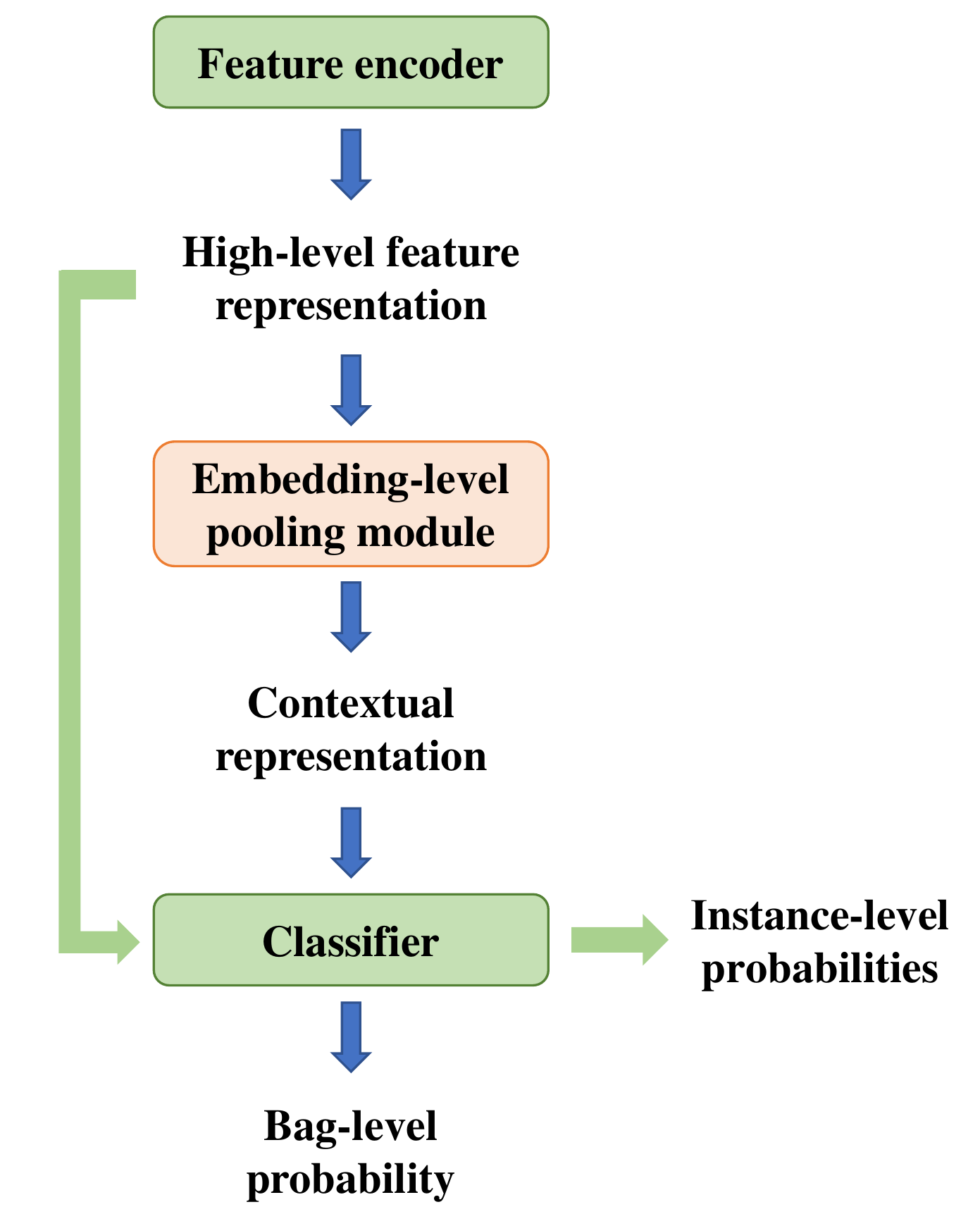}
  \caption{The instance-level probabilities for the embedding-level approach.}
  \label{fig0}
	\vskip -0.1in
\end{figure}

According to Equation~\ref{eq11}$, \mathbf{h}_c$ in GMP only relates to two frames in an audio clip, the value of the high-level feature representation of which in one dimension is the largest of all frames, as shown in Figure~\ref{fig2a}. Thus the movements in GMP affect fewer frames in a negative audio clip but make fewer mistakes in a positive audio clip (carry fewer false-positive frames toward the positive side of the decision surface). Similarly, $\mathbf{h}_c$ in GAP relates to all the frames and focuses on them equally as shown in Figure~\ref{fig2b}, for which it affects all the frames in a negative audio clip but makes more mistakes in a positive audio clip. For GSP, according to Equation~\ref{eq9} and~\ref{eq13}, the strength of the connection $a_{ct}$ depends on $\hat{p}(y_c\mid \mathbf{x}_t)$. Since we consider that $\mathbf{x}_t$ and $\mathbf{h}_c$ share the same decision surface, the strength of the connection $a_{ct}$ in GSP exactly depends on this shared decision surface. Therefore, the strength of the connection between $\mathbf{x}_t$ and $\mathbf{h}_c$ exactly depends on how much the model considers it as a positive frame. As shown in Figure~\ref{fig2c}, where the black solid line represents the decision surface mentioned above, if we ignore some of the relatively weak connection, those green lines between $\mathbf{h}_c$ and $\mathbf{x}_t$ lying on the negative side of the decision surface can be neglected. Hence, GSP pursues a trade-off between affecting more frames in a negative audio clip and making fewer mistakes in a positive audio clip.

%%The movements in GMP affect fewer frames once a time but make fewer mistakes, while those in GAP are exactly the opposite. Differently, GSP pursues a trade-off. Attribute to these movements, positive frames and negative frames are able to gradually separate into two clusters and the decision surface of the contextual representations is close to the boundary of such two clusters, in other words, suitable for coarse frame-level classification, for which we regard it as a shared decision surface.

Therefore, for the embedding-level approach, for GMP, GAP and GSP, as shown in Figure~\ref{fig0}, we pass the frame-level high-level feature representation $\mathbf{x}_t$ through clip-level classifier to get frame-level probabilities $\hat{p}\left(y_c\mid \mathbf{x}_t\right)$ for event $c$ at time $t$. Then the the frame-level prediction is:
\begin{equation}
\varphi_c\left(\mathbf{x},t\right)
=\left\{\begin{matrix}
1,&G_c(\mathbf{x}_t)\cdot\phi_{c}\left(\mathbf{x}\right)\geq \gamma \\ 
0,&{\rm otherwise}
\end{matrix}\right.
,
\label{eq3}
\end{equation}
where $G_c$ denotes the classifier of the contextual representation $\mathbf{h}_c$ according to Equation~\ref{eq5}.

\subsection{Specialized decision surface}
\label{Specialized decision surface}

To explore the more accurate boundary of the two clusters mentioned above, we propose a specialized decision surface (SDS). Different from the shared decision surface, SDS is not approximately close to but exactly the boundary of the two clusters so that SDS is able to provide more accurate frame-level detection.

Actually, for GMP and GAP, SDS does not present in an explicit way. Intuitively, they do not provide a explicit way to separate these two clusters. However, as for GSP, according to Equation~\ref{eq8} and \ref{eq9}, the forming of the two clusters depends on $a_{ct}$ relating to the shared decision surface $G_c$, so that SDS of GSP is coincident with the shared decision surface of GSP.

When it comes to ATP, according to Equation~\ref{eq10}, the strength of the connection $a_{ct}$ depends on the independent detector discussed in Section~\ref{Multiple instance learning for polyphonic SED} instead of the shared decision surface $G_c$ as shown in Figure~\ref{fig2d}. Therefore, free parameters $\mathbf{w}_c$ in the independent detector (dotted line in Figure~\ref{fig2d}) determine how to chose frames to move with the contextual representation together and gradually separate these frames from the rest. Although these movements try to promote the two clusters to distribute separately on opposite sides of the shared decision surface, the SDS utilized directly to select and separate the two clusters can better match the boundaries of the two clusters.

%as shown in Figure~!!!!,
Therefore, the implement of SDS for the embedding-level attention pooling is based on a frame-level classifier $S_c$: the combination of the independent detector utilized to generate $a_{ct}$ and an activation layer employed to generate probabilities.

%On the basis of this observation and the discussion in the last section, we argue that the exact boundary (dotted line in Figure~\ref{fig2d}) of the two clusters mentioned above provides a superior decision surface, which we term as SDS, to the shared decision surface in frame-level detection.

%Similarly, SDS of ATP depends on free parameters in the independent detector (dotted line in \ref{fig2d}). Compared with GSP, ATP with SDS no longer makes $\mathbf{h}$ and $\mathbf{x}$ share the same decision surface so that it allows more flexible forming of $\mathbf{x}$ and leads more accurate frame-level detection.

Then the frame-level prediction for event $c$ at time $t$ is:
\begin{equation}
\varphi_c\left(\mathbf{x},t\right)
=\left\{\begin{matrix}
1,&S_c(\mathbf{x}_t)\cdot\phi_{c}\left(\mathbf{x}\right)\geq \gamma \\ 
0,&{\rm otherwise}
\end{matrix}\right.
,
\label{eq14}
\end{equation}

\begin{equation}
S_c\left(\mathbf{x}_t\right)=
\sigma\left( \mathbf{w}_c^{\mathsf{T}} \mathbf{x}_t\right),
\label{eq15}
\end{equation}
where $\mathbf{w}_c$ are free parameters of the independent detector and $\sigma$ is an activation function to generate probabilities. We take Sigmoid as this activation function in our work.

\subsection{Disentangled feature}
\label{Disentangled feature}
For all the MIL approaches described above, the general feature encoder generates the high-level feature representations of all the categories from the same feature space. However, for multi-label classification, when a certain category often appears in co-occurrence with other categories, this approach makes it difficult to differentiate every single category. In other words, the forming of the high-level feature subspace of the event categories with insufficient identifiable information given in the training set will be largely disturbed by those categories appearing in co-occurrence with them. This effect will be exacerbated when the number of clips with much identifiable information of certain categories in the unbalanced set is particularly small.

To mitigate this effect, we propose DF to re-model multiple feature subspaces for multiple categories. In this way, every category shares a different part of the feature encoder instead of the whole feature encoder and is allocated in advance a feature subspace of the high-level feature space generated by the feature encoder according to its priori information.

%Since $\mathbf{h}$ is produced by $\mathbf{x}$ according to Equation~\ref{eq8}, the feature space of $\mathbf{h}$ is also re-modeled into $C$ feature subspaces.

Assuming that ${\chi}^d$ ($\mathbf{x}\in {\chi}^d$) is a $d$-dimensional space generated by the feature encoder and $\mathbf{B} =\left\{\mathbf{b}_1,\mathbf{b}_2,\ldots,\mathbf{b}_d\right\}$ is a basis of ${\chi}^d$. We define $\chi_c$, a subspace of ${\chi}^d$, as the feature space of event category $c$. We produce $\chi_c$ by selecting specific base vectors from $\mathbf{B}$ and the basis of $\chi_c$ is
\begin{equation}
\mathbf{B}_c =\left\{\mathbf{b}_1,\mathbf{b}_2,\ldots,\mathbf{b}_{k_c}\right\},
\end{equation}
where $\mathbf{k}=\left\{k_1,k_2,\ldots,k_C\right\} \left(0<k_c \leqslant d\right)$ relates to the volume of $\chi_c$.

\begin{figure}[t]
 \centering
  \includegraphics[width=\columnwidth]{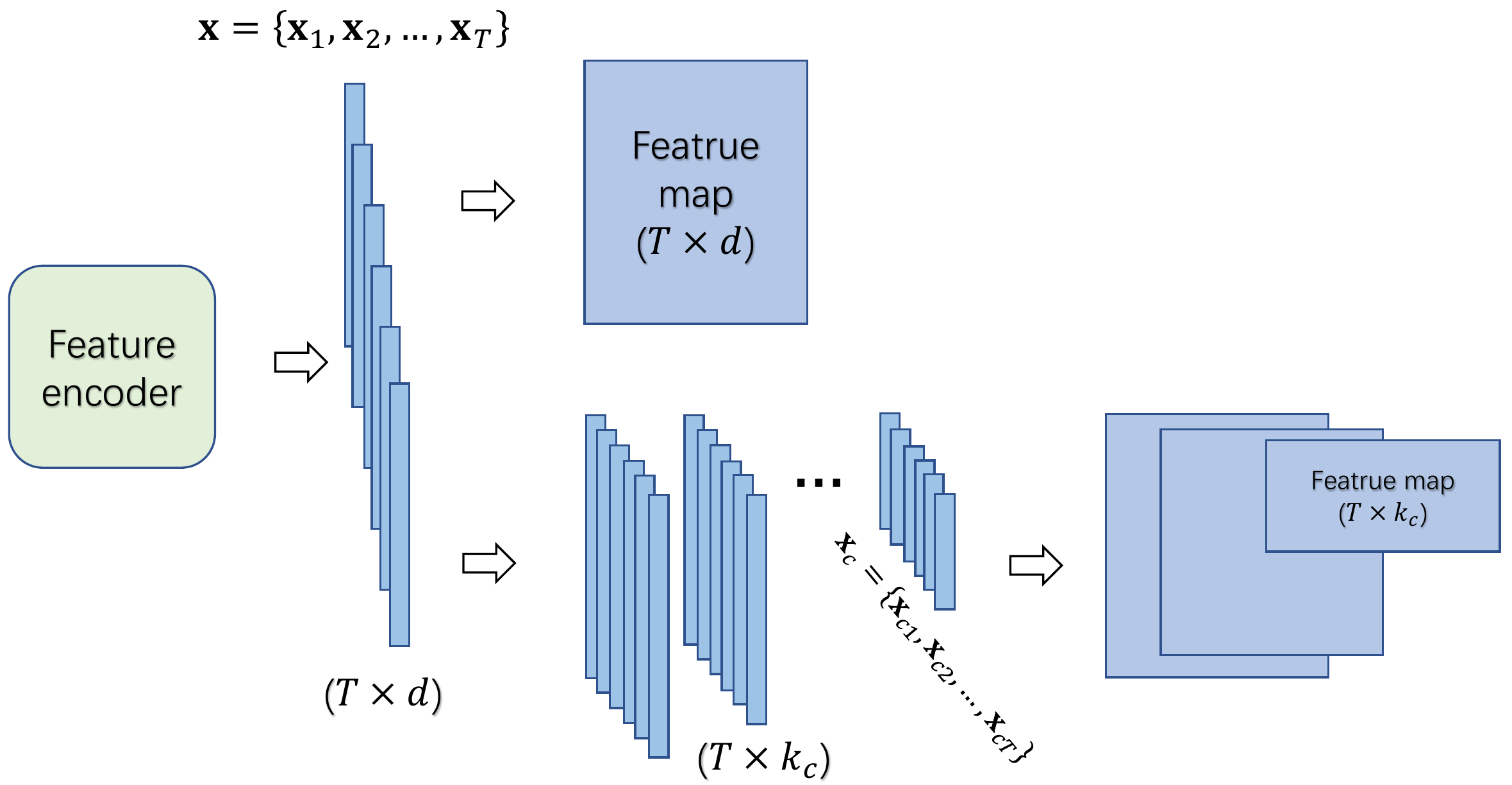}
  \caption{The comparison of general feature (top) and disentangled feature (bottom).}
  \label{fig3}
\end{figure}
In this way, the diversity of elements in $\mathbf{k}$ leads to the feature space of each category being remodeled into a disentangled feature space that is different from those of the other categories. 
For two categories $i$ and $j$, the larger the absolute value of the difference between $k_i$ and $k_j$ is, the more different their feature space will be. The difference of feature spaces results in the diversity of decision surfaces among different categories without pre-training. In the extreme case with $k_1=k_2=\ldots=k_C=d$, all subspaces are equal to ${\chi}^d$ so that the disentangled feature degenerates to general feature.

Meanwhile, if a clip contains more categories, we argue that for each of these categories, the clip contains more interference from other categories. We argue that the volume of $\chi_c$ is determined by the number of available clips containing little interference. This is because that for category $c$, the larger the proportion of the clips containing little interference from other event categories is, the more the class-wise identifiable information needs to be learned, which requires the larger volume of the feature space. In contrast, the smaller the proportion of these clips is, the smaller volume of the feature space is required to prevent overfitting. For this reason, $k_c$ increases as the proportion of these clips of category $c$ increases. 

Considering that too-small $k_c$ severely cut into the ability of the model to recognize category $c$, we utilize a constant factor $m \left(0\leqslant m \leqslant 1\right)$ to tackle this effect, then,
\begin{equation}
k_c =\lceil\left(\left(1-m\right)\cdot f_c+m\right)\cdot d\rceil,
\end{equation}
where $f_c \left(0\leqslant f_c \leqslant 1\right)$ relates to the number of clips containing little interference in the training set. As $m$ increases to $1$, DF degrades into the general feature.

We quantify the level of interference according to the principle that the more categories a clip covers, the more interference the other categories cause to any one of them, then,
\begin{equation}
f_c = \sum_{i=1}^{C}\frac{r_i\cdot {N_c}_i}{R},
\end{equation}
\begin{equation}
R = \underset{c}{\max}\; \sum_{i=1}^{C}r_i\cdot {N_c}_i.
\end{equation}

Here, ${N_c}_i$ denotes the number of clips containing $i$ categories including category $c$ in the training set and $r_i$ is corresponding constant coefficient implying the importance of these clips. We argue that the less interference the other categories cause to any one of them in a clip, the more important the clip is, for which we determine $r_i$ as:
\begin{equation}
 r_i=\frac{1}{i} \left(1\leqslant i \leqslant C\right).
 \label{eq20}
\end{equation}

We can also just consider those clips containing the least interference, then,
\begin{equation}
r_i=\left\{\begin{matrix}
 1,& i=1\\ 
 0,& {\rm otherwise}
\end{matrix}\right..
\label{eq21}
\end{equation}

To simplify training, we take an orthogonal basis 
$\mathbf{B^{'}} =\left\{\mathbf{e}_1,\mathbf{e}_2,\ldots,\mathbf{e}_d\right\}$ where the element of $\mathbf{e}_i$ in the $i^{th}$ dimension is 1 for ${\chi}^d$. Then the $k_c$ basis vectors are related to $k_c$ dimensions of $\mathbf{x}_t$. As shown in Figure~\ref{fig3}, we easily get a ladder-shape group of disentangled feature maps from feature encoder for a clip.

Combining disentangled feature $\mathbf{x}_c=\left\{\mathbf{x}_{c1},\mathbf{x}_{c2},\ldots,\mathbf{x}_{cT}\right\}$ and the embedding-level attention module, to generate the contextual representation of event category $c$, we have
\begin{equation}
\hat{P}\left(y_c\mid\mathbf{x}\right)=\hat{P}\left(y_c\mid \mathbf{x}_c\right)=\hat{P}\left(y_c\mid \mathbf{h}_c\right),
\end{equation}
\begin{equation}
  \mathbf{h}_c=\sum_t a_{ct}\cdot \mathbf{x}_{ct}.
\end{equation}

Then the contribution of $\mathbf{x}_{ct}$ to an audio clip is:
\begin{equation}
a_{ct}=\frac{\exp\left( \mathbf{w}_c^{\mathsf{T}}\mathbf{x}_{ct}/d_c\right)}{\sum_k \exp\left(\mathbf{w}_c^{\mathsf{T}}\mathbf{x}_{ck}/d_c\right)},
\end{equation}
where $ \mathbf{w}_c^{\mathsf{T}}$ are learnable parameters mentioned in Section~\ref{Instance-level pooling modules} and $d_c$ is a scaling factor consistent with the dimensions of $\mathbf{x}_{ct}$.

%\begin{figure}[t]
%  \centering
%  \includegraphics[width=\columnwidt%h]{DCASE2018_data.png}
%  \caption{ asasas.}
%  \label{fig4}
% \end{figure}
 
\section{Experiments}
\label{Experiments}
In this section, we introduce the dataset and describe in detail the model architecture, the pre-processing, and post-processing methods, the training configuration, and the evaluation measure in our experiments. 

\begin{figure}[t]
  \centering
  \includegraphics[width=0.5\columnwidth]{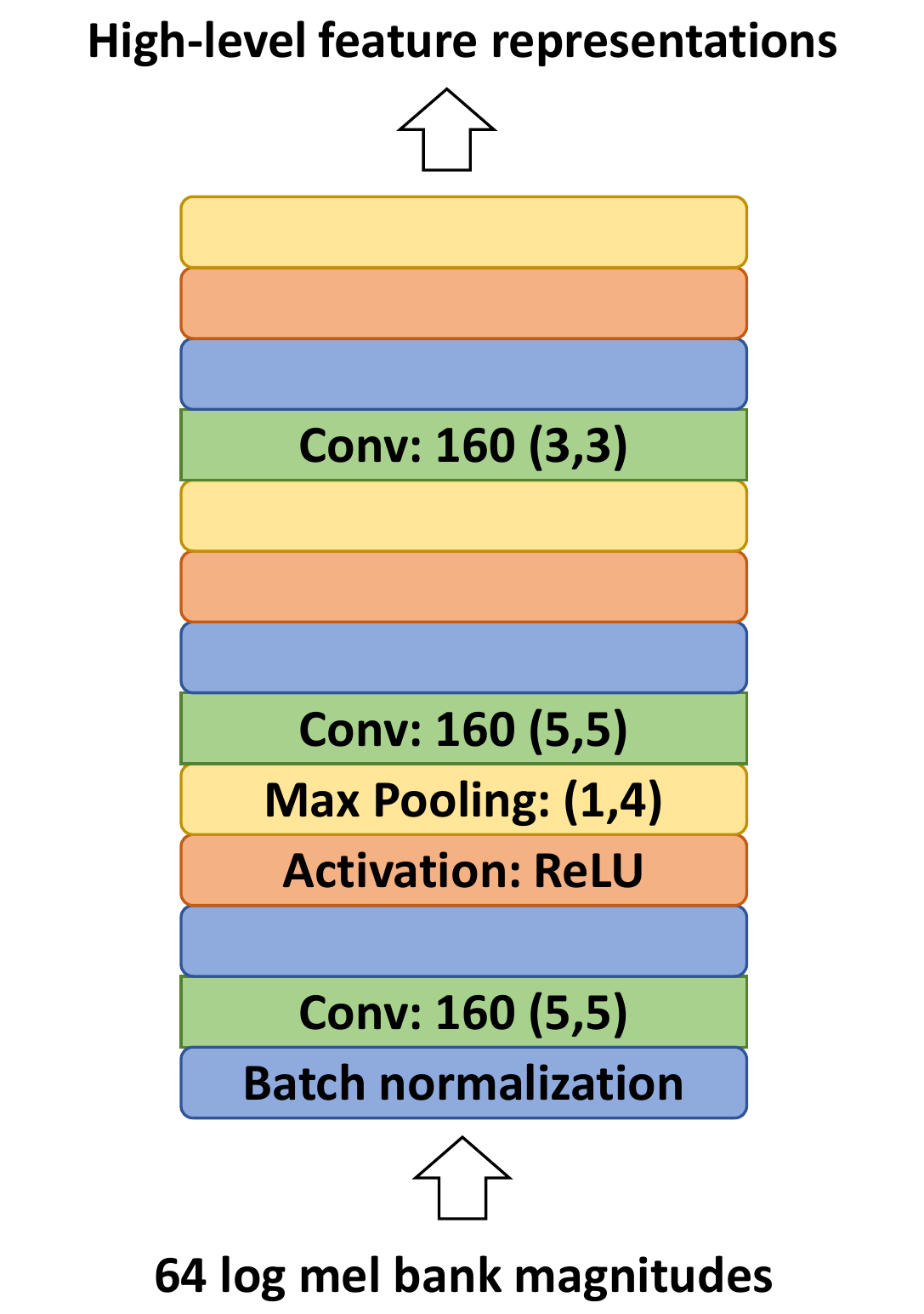}
  \caption{The architecture of the feature encoder.}
  \label{fig5}
\end{figure}

\subsection{Dataset}
We utilize the dataset from task 4 of the DCASE 2018 Challenge \cite{serizel2018_DCASE}, which is a subset of AudioSet \cite{gemmeke2017audio} by Google. The dataset consists of 10 categories of sound events from domestic environment: alarm/bell/ringing, blender, cat, dishes, dog, electric shaver/toothbrush, frying, running water, speech, and vacuum cleaner. The set contains 1578 weakly-labeled clips (2244 event occurrences) for which weak annotations have been verified and cross-checked, 14412 unlabeled in domain clips, 39999 unlabeled out-of-domain clips and 1168 clips with strong annotations. The challenge divides strong-labeled clips into two subsets: a validation set (288 clips) and an evaluation set (880 clips). In our experiments, we utilized the weakly labeled data to pre-train a clip-level classification model to tag unlabeled in domain data with weak annotations and wipe off 1001 clips with empty annotations.
The clip-level classification model achieves a micro F-measure of 0.688 and a macro F-measure of 0.627 on audio tagging on the test set. Since such the feature encoder of such a model only outputs 5 frame-level high-level feature representations, the event detection performance of it is relatively poor. But it is suitable for tagging unlabeled data with weak annotations. We combine these noisy-annotated data with the original weakly-labeled training set because the weakly-labeled training set is too small to achieve stable performances of different pooling methods. Since there are also some errors occurring in annotations of some large-scale manual-labeled dataset, we argue that such a combination of data set makes sense. Consequently, the training set in our experiments embraces 14989 clips with noisy weak annotations, the characteristic of which are large scale and unbalanced distribution. Besides, there are other problems contained in this dataset, such as noisy annotations, the co-occurrence of multiple categories, overlapping sounds and the appearances of the sound events outside of the target categories.

\subsection{Model architecture}
The models employed in our experiments are divided into the instance-level model and the embedding-level model. As shown in Figure~\ref{figp1}, both these two types of models comprise three modules: the feature encoder, the pooling module, and the classifier. The feature encoder is designed based on the model architecture of the baseline system of the task 4 \cite{serizel2018_DCASE}. We remove the RNN layer to make each frame-level high-level feature representation contain more identifiable information about the current frame, for which finer frame-level information is maintained. Meanwhile, the model thus depends more on the pooling module to integrate contextual information. We also remove dropout layers and increase the number of filters of CNN layers. The final feature encoder consists of 3 convolutional blocks, each of which comprises a convolutional layer, a batch normalization \cite{ioffe2015batch} layer, a max pooling layer (no temporal pooling), and an activation layer, as shown in Figure~\ref{fig5}. The pooling modules including GAP, GMP, GSP and ATP are described in detail in Section~\ref{Instance-level pooling modules} and Section~\ref{Embedding-level pooling modules}. We utilizes $1\times 1$ convolutional layer with Sigmoid activation function as the classifier.

Different from other general pooling modules in the prediction phase, the instance-level and embedding-level ATP-SDS make frame-level prediction according to Equation~\ref{eq14} and Equation~\ref{eq15} discussed in Section~\ref{Specialized decision surface}.

\begin{figure}[t]
  \centering
  \includegraphics[width=\columnwidth]{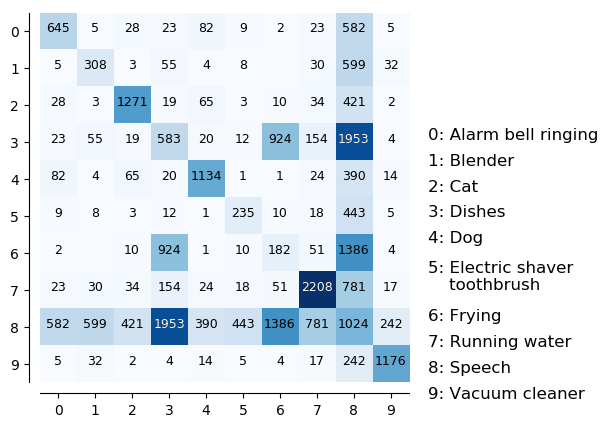}
  \caption{The number of clips where two categories occur in co-occurrence. The diagonal represents the number of the clips containing only one category.}
  \label{fig4}
\end{figure}

\begin{figure}[t]
  \centering
  \includegraphics[width=0.82\columnwidth]{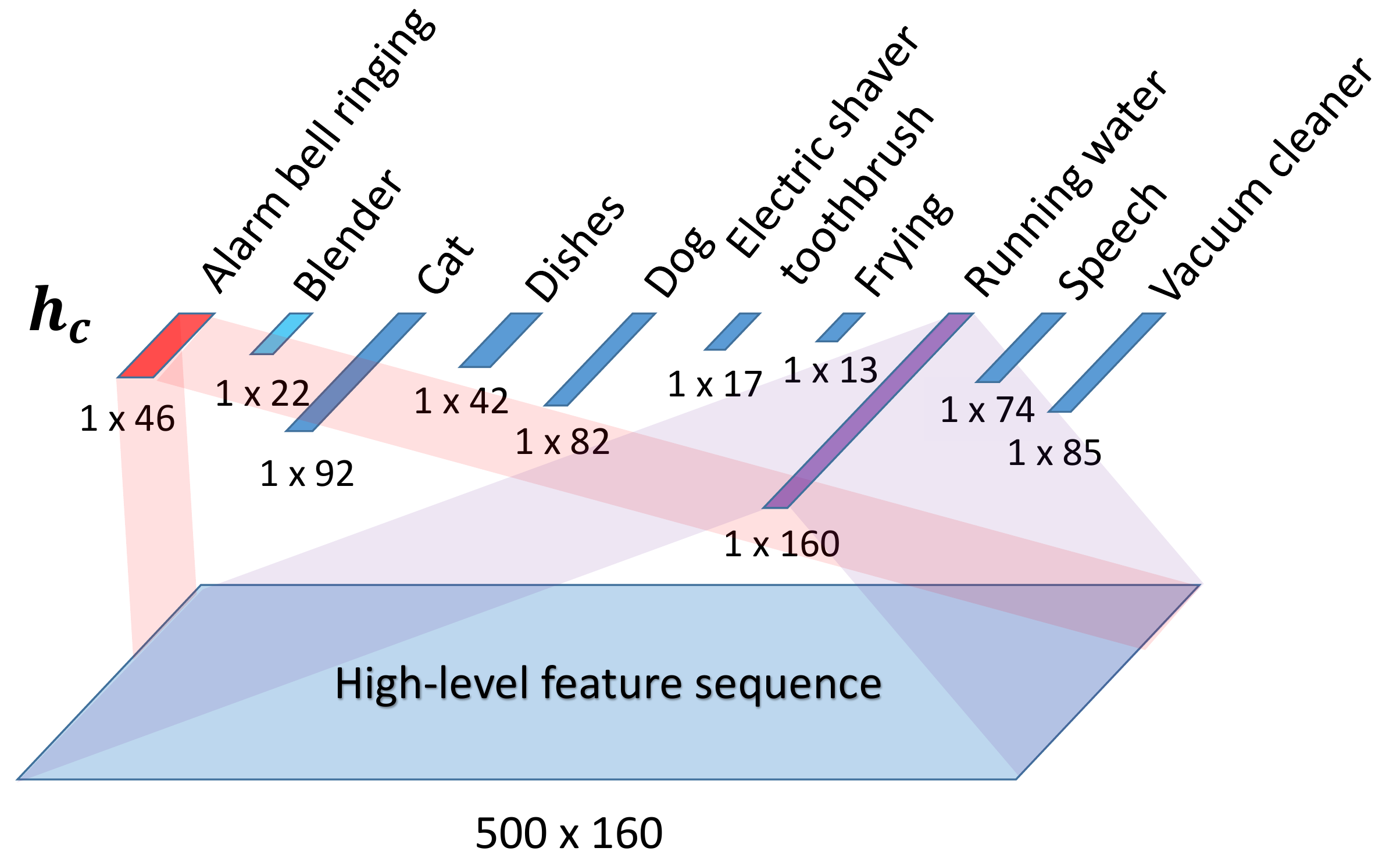}
  \caption{A sketch of the high-level feature representations of the the embedding-level ATP-SDS with DF1.}
  \label{fig6}
\end{figure}

As for DF, we experiment with two different methods for determination of the constant coefficient $r_i$ discussed in Section~\ref{Disentangled feature}: the embedding-level ATP-SDS with DFW (Equation~\ref{eq20}) and the embedding-level ATP-SDS with DF1 (Equation~\ref{eq21}).
In addition, as mentioned in Section~\ref{Disentangled feature}, since the hyper-parameter $m$ is to avoid too-small $k_c$ and for this dataset, each $k_c$ is within a reasonable range, we set $m=0$ in our experiments. Figure~\ref{fig4} illustrates the details of the co-occurrence of every two sound event categories on the training set and Figure~\ref{fig6} illustrates the condition of the embedding-level ATP-SDS with DF1. More detailed information of disentangled dimessions for each category of the DFW and DF1 methods is shown in Table~\ref{table1}.


\begin{figure}[t]
  \centering
  \includegraphics[width=0.98\columnwidth]{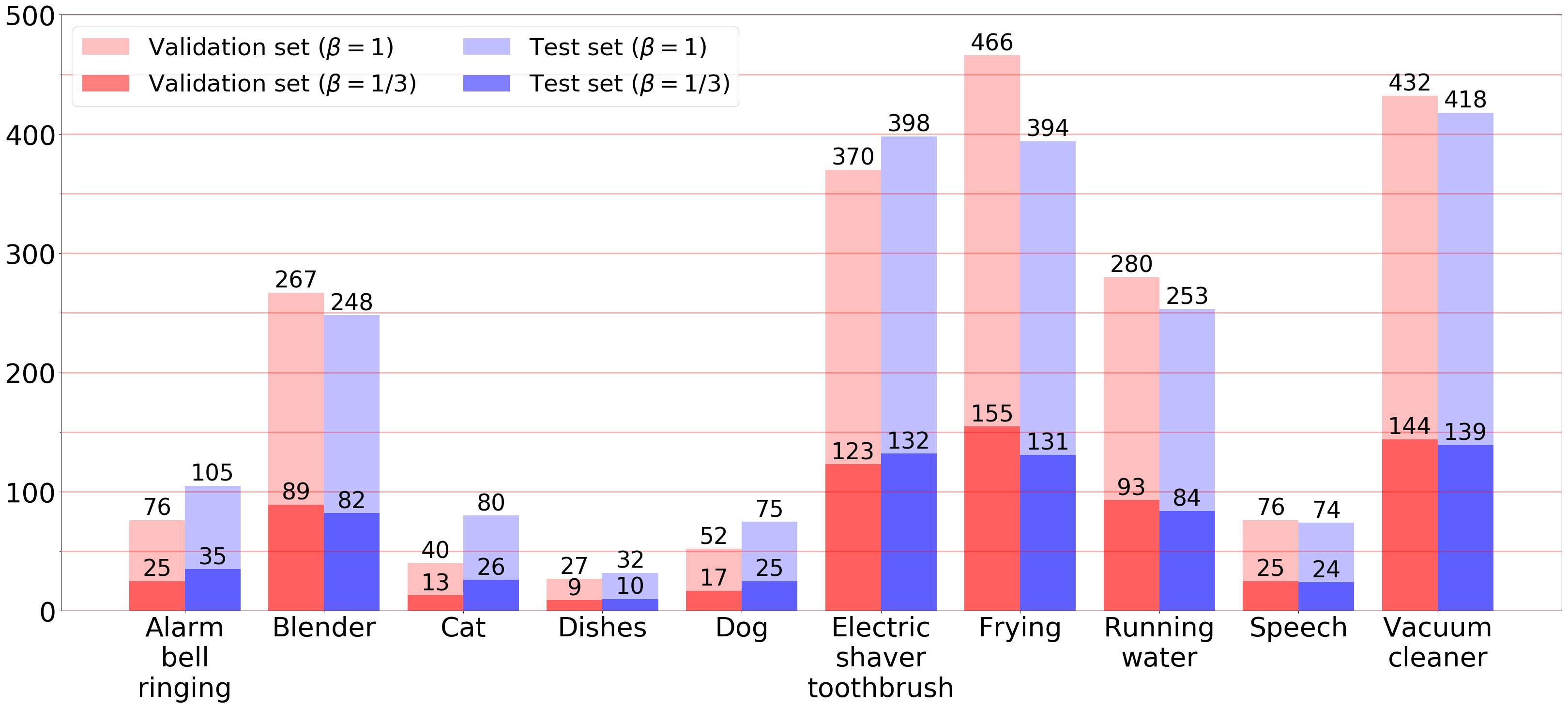}
  \caption{The the window sizes calculated from the validation set and the test set with $\beta = 1$ and $\beta = \frac{1}{3}$ respectively.}
  \label{fig12}
\end{figure}

\begin{table}[t]
  \caption{The DF dimension and the window size of median filters when $\beta=\frac{1}{3}$ per category.}
  \label{table1}
  \centering
  \setlength{
\tabcolsep}{2mm}{
\begin{tabular}{c|c|c|c}
\hline
\multirow{2}{*}{\textbf{Event}}&\multicolumn{2}{c|}{\textbf{DF dimension}}&\multirow{2}{*}{\textbf{Window Size (frame)}}\\
\cline{2-3}
&DF1&DFW\\
\hline
Alarm bell ringing&
46&31&17\\
\hline
Blender&22&22&42\\
\hline
Cat&92&43&17\\
\hline
Dishes&42&66&9\\
\hline
Dog&82&39&16\\
\hline
Electric shaver toothbrush&17&16&74\\
\hline
Frying&13&41&85\\
\hline
Running water&160&75&64\\
\hline
Speech&74&160&18\\
\hline
Vacuum cleaner&85&35&87\\
\hline
\end{tabular}}
\end{table}

\subsection{Pre-processing and post-processing}
The feature passed into the feature encoder employed 64 log mel-bank magnitudes (sampling rate 44.1 KHz) which are extracted from 40 ms frames with $50\%$ overlap $(n_{\mathrm{FFT}}=2048)$ using the librosa package \cite{brian_mcfee-proc-scipy-2015}. All the 10-second audio clips are transformed to feature vectors with 500 frames. The threshold $\alpha$ (mentioned in Section~\ref{Multiple instance learning for polyphonic SED}) of the predicted probability to determine whether an event category exists in a clip is $0.5$. For frame-level prediction, all the probabilities are smoothed by a median filter with a group of adaptive window sizes. The operation of smoothing is repeated on the final frame-level prediction with a threshold $\gamma = 0.5$.

\begin{table*}[t]
\renewcommand{\arraystretch}{1.4}
\caption{The average performance of models.}
\label{table3}
\centering
\setlength{
\tabcolsep}{2mm}{
\begin{tabular}{llcccccccc}
\hline
&&&
\multicolumn{3}{c}{\textbf{Event detection (frame-level)}}&&
\multicolumn{3}{c}{\textbf{Audio tagging (clip-level)}}\\
\hline
\multicolumn{2}{c}{\textbf{Model}} && $\mathbf{F_1}$&$\mathbf{P}$&$\mathbf{R}$&&
$\mathbf{F_1}$&$\mathbf{P}$&$\mathbf{R}$\\
\cline{1-2}\cline{4-6}\cline{8-10}

\multirow{5}{*}{\shortstack{\textbf{Instance}\\\textbf{-level}\\
\textbf{pooling}}}&\textbf{GMP} &
& $0.013\pm0.002$
& $0.089\pm0.018$
& $0.007\pm0.001$
&
& $0.565\pm0.007$
& $0.643\pm0.014$
& $0.531\pm0.015$\\
&\textbf{GAP} &
& $0.187\pm0.004$
& $0.248\pm0.007$
& $0.172\pm0.003$
&
& $0.471\pm0.003$
& $0.668\pm0.003$
& $0.403\pm0.004$\\
&\textbf{GSP} &
& $0.208\pm0.006$
& $0.266\pm0.006$
& $0.190\pm0.006$
&
& $0.487\pm0.006$
& $0.682\pm0.007$
& $0.412\pm0.005$\\

&\textbf{ATP} &
& $0.260\pm0.003$
& $0.298\pm0.007$
& $0.255\pm0.005$
&
& \multirow{2}{*}{\shortstack{$0.612\pm0.010$}}
& \multirow{2}{*}{\shortstack{$\mathbf{0.743\pm0.009}$}}
& \multirow{2}{*}{\shortstack{$0.555\pm0.011$}}\\
&\textbf{ATP-SDS} &
& $0.284\pm0.007$
& $0.347\pm0.006$
& $0.271\pm0.007$
&
& 
& 
& \\
\hline
\multirow{5}{*}{\shortstack{\textbf{Embedding}\\\textbf{-level}\\
\textbf{pooling}}}&\textbf{GMP} &
& $0.239\pm0.006$
& $0.249\pm0.009$
& $0.245\pm0.006$
&
& $0.626\pm0.005$
& $0.674\pm0.006$
& $0.610\pm0.008$\\

&\textbf{GAP} &
& $0.242\pm0.004$
& $0.247\pm0.006$
& $0.265\pm0.005$
&
& $0.602\pm0.005$
& $0.657\pm0.005$
& $0.597\pm0.007$\\

&\textbf{GSP} &
& $0.242\pm0.004$
& $0.243\pm0.006$
& $0.268\pm0.005$
&
& $0.608\pm0.005$
& $0.660\pm0.006$
& $0.603\pm0.007$\\
&\textbf{ATP} &
& $0.253\pm0.004$
& $0.247\pm0.005$
& $0.301\pm0.007$
&
& \multirow{2}{*}{\shortstack{$0.620\pm0.005$}}
& \multirow{2}{*}{\shortstack{$0.665\pm0.005$}}
& \multirow{2}{*}{\shortstack{$\mathbf{0.634\pm0.007}$}}\\
&\textbf{ATP-SDS} &
& $0.349\pm0.004$
& $0.364\pm0.007$
& $0.370\pm0.003$
&
& 
& 
& \\
\hline

\multirow{2}{*}{\shortstack{
\textbf{eATP-SDS}}}
&\textbf{DFW} &
& $0.358\pm0.004$
& $0.365\pm0.007$
& $\mathbf{0.374\pm0.005}$
&
& $\mathbf{0.639\pm0.006}$
& $0.682\pm0.006$
& $0.624\pm0.009$\\
&\textbf{DF1} &
& $\mathbf{0.359\pm0.008}$
& $\mathbf{0.373\pm0.008}$
& $0.373\pm0.006$
&
& $0.634\pm0.006$
& $0.679\pm0.005$
& $0.620\pm0.006$\\

\hline
\end{tabular}}
\end{table*}

The adaptive window size of the median filter for category $c$ is:
\begin{equation}
{\mathrm{win}}_c={\mathrm{duration}}_{c}\cdot\beta,
\end{equation}
where ${\mathrm{duration}}_{c}$ is the average duration of category $c$ in the validation set. In addition, we set $\beta=\frac{1}{3}$ and shows the specific window sizes in Table~\ref{table1}. 

We argue that the average duration of each event is more like a fixed physical property. In real life, the duration of the same event is similar. Therefore, we argue the duration of the validation set is representative enough. As shown in Figure~\ref{fig12}, we show the window sizes calculated from the validation set and the test set, and their distribution is similar. Since the model early stops according to the audio tagging performance of the validation set rather than the detection performance and the audio tagging performance has nothing to do with the post-processing method, the duration calculated from the validation set will not make the selected model heavily overfit the validation set. Besides, we argue that such an adaptive method tends to be superior to the method with fixed window sizes. And it is also more convenient and reliable than using empirical value to decide window sizes.

\begin{table}[t]
\renewcommand{\arraystretch}{1.4}
\caption{The performance of the models with the best frame-level $F_1$ score in $20$ experiments.}
\label{table4}
\centering
\setlength{
\tabcolsep}{1mm}{
\begin{tabular}{llcccccc}
\hline
&&
\multicolumn{3}{c}{\textbf{Event detection}}&
\multicolumn{3}{c}{\textbf{Audio tagging}}\\
\hline
\multicolumn{2}{c}{\textbf{Model}} & $\mathbf{F_1}$&$\mathbf{P}$&$\mathbf{R}$&
$\mathbf{F_1}$&$\mathbf{P}$&$\mathbf{R}$\\
\hline
\multicolumn{2}{c}{\textbf{Baseline system \cite{serizel2018_DCASE}}}&0.141&-&-&-&-&-\\
\multicolumn{2}{c}{\textbf{The 1$^{st}$ place \cite{Lu2018}}}&0.324&-&-&-&-\\
\hline
\multirow{5}{*}{\shortstack{\textbf{Instance}\\\textbf{-level}\\
\textbf{pooling}}}&\textbf{GMP}
& $0.023$
& $0.094$
& $0.014$

& $0.596$
& $0.641$
& $0.576$\\
&\textbf{GAP} 
& $0.202$
& $0.271$
& $0.186$

& $0.482$
& $0.670$
& $0.410$\\
&\textbf{GSP} 
& $0.240$
& $0.302$
& $0.218$

& $0.518$
& $0.696$
& $0.426$\\
&\textbf{ATP} 
& $0.273$
& $0.315$
& $0.260$

& \multirow{2}{*}{\shortstack{$0.636$}}
& \multirow{2}{*}{\shortstack{$\mathbf{0.735}$}}
& \multirow{2}{*}{\shortstack{$0.586$}}\\

&\textbf{ATP-SDS} 
& $0.312$
& $0.364$
& $0.302$

& 
& 
& \\
\hline
\multirow{5}{*}{\shortstack{\textbf{Embedding}\\\textbf{-level}\\
\textbf{pooling}}}&\textbf{GMP}
& $0.271$
& $0.274$
& $0.286$

& $0.648$
& $0.683$
& $0.639$\\

&\textbf{GAP} 
& $0.258$
& $0.266$
& $0.278$

& $0.614$
& $0.656$
& $0.598$\\
&\textbf{GSP} 
& $0.252$
& $0.245$
& $0.279$

& $0.628$
& $0.667$
& $0.616$\\
&\textbf{ATP} 
& $0.267$
& $0.268$
& $0.309$

& \multirow{2}{*}{\shortstack{$0.643$}}
& \multirow{2}{*}{\shortstack{$0.663$}}
& \multirow{2}{*}{\shortstack{$\mathbf{0.663}$}}\\

&\textbf{ATP-SDS} 
& $0.362$
& $0.398$
& $0.365$

& 
& 
& \\
\hline
\multirow{2}{*}{\shortstack{\textbf{eATP-SDS}}}
&\textbf{DFW} 
& $0.380$
& $0.397$
& $0.375$

& $\mathbf{0.658}$
& $0.676$
& $0.649$ \\   
&\textbf{DF1} 
& $\mathbf{0.390}$
& $\mathbf{0.402}$
& $\mathbf{0.402}$

& $0.650$
& $0.684$
& $0.632$ \\

\hline
\end{tabular}}
\end{table}

\subsection{Training and evaluation}
The neural networks are trained using the Adam optimizer \cite{kingma2014adam} with learning rate of $0.0018$ and mini-batch of $64$ 10-second patches. The learning rate is reduced by $20\%$ per $10$ epochs. We take binary cross entropy as loss function. Training stops if there is no more improvement in clip-level macro $F_1$ performance on the validation set within $10$ epochs. The best performing model on the validation set will be retained for prediction before the training stops. All the experiments are repeated $20$ times under the same parameter configuration. We fix random seeds of the CPU and the GPU, but some operations on the GPU have non-deterministic outputs. Therefore, the results of these 20 experiments were not exactly the same. We took the average of all the results as the final result. In particular, in order to compare with the performance of the first place in the challenge, we report the best results among these $20$ experiments in addition. For event detection, event-based measures (macro-averaged) \cite{Mesaros2016_MDPI} with a 200 ms collar on onsets and a 200 ms / $20\%$ of the events length collar on offsets are calculated over the entire test set. For audio tagging, we report the macro $F_1$-score. The implementation of our methods is available online at https://github.com/Kikyo-16/Sound\_event\_detection.

\section{discussion}
\label{discussion}
In this section, we report the results of our experiments and analyze in detail the distribution of data in the high-level feature space of models to prove our conjecture.

\begin{figure}[t]
  \centering
  \includegraphics[width=0.81\columnwidth]{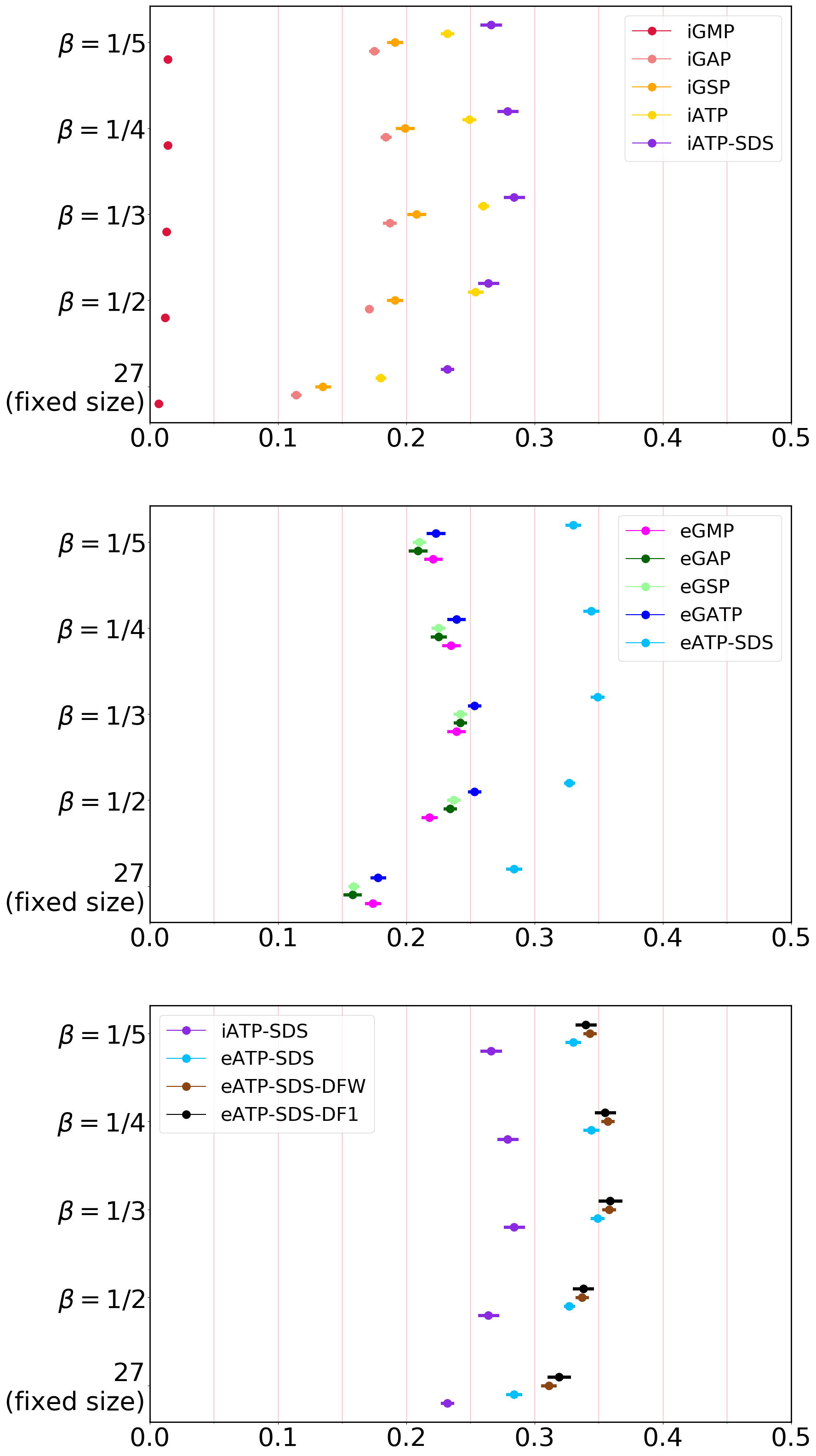}
  \caption{The frame-level $F_1$ score of all $20$ experiments of all the models with different window sizes of median filters. i* such as iGMP represents the instance-level model and e* such as eGMP represents the embedding-level model.}
  \label{fig7}
  \vskip -0.2in
\end{figure}

\begin{figure*}[t]
\vskip -0.1in
  \centering

    \includegraphics[width=1.6\columnwidth]{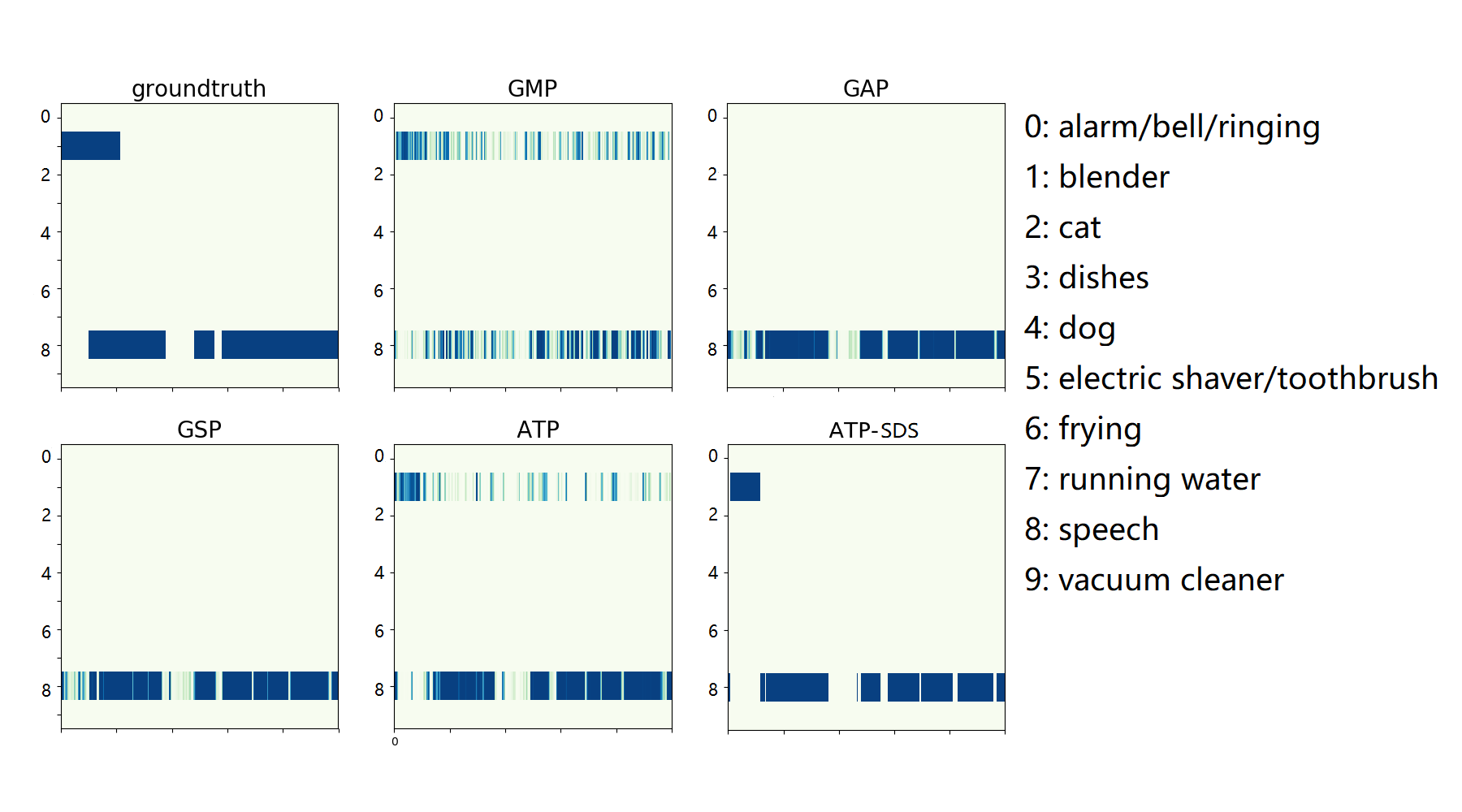}
    \vskip -0.2in
  \caption{Comparison of frame-level possibilities output with the groundtruth. The example is cherry-picked to indicate the points we highlight.}
  \label{fig9}

\end{figure*}
\begin{figure*}[t]
  \centering
  \includegraphics[width=1.8\columnwidth]{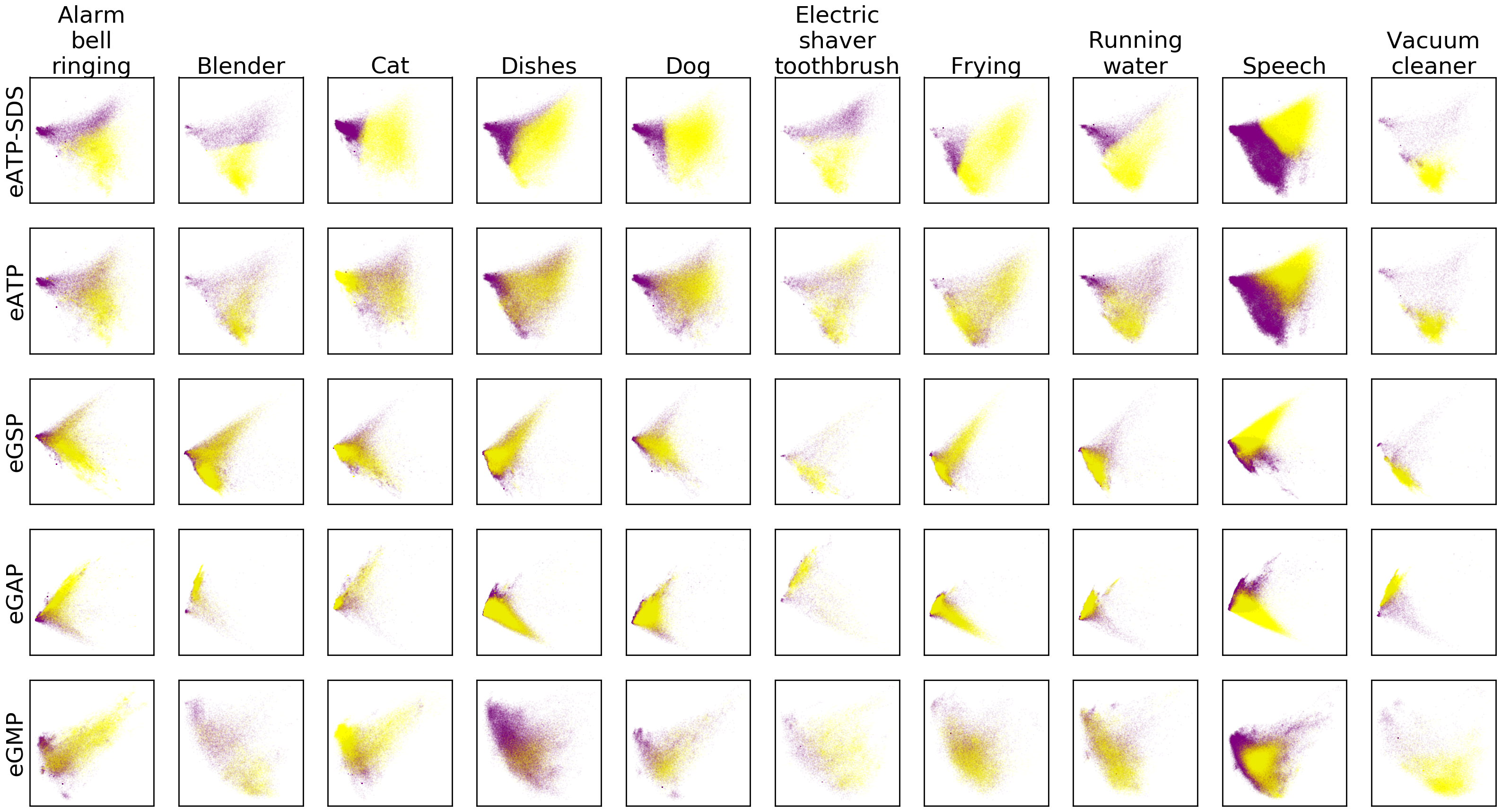}
  \caption{The decision surfaces for different categories in the feature space generated from feature encoder (PCA).}
  \label{fig10}
\end{figure*}
\subsection{Results}
We report the average results (with the $95\%$ confidence intervals) of 20 experiments of all the models in Table~\ref{table3} and the performance of the models with the best frame-level F1 score among these 20 experiments in Table~\ref{table4}. As shown in Table~\ref{table3}, the embedding-level ATP-SDS with DF1 achieves the best average frame-level $F_1$ score of $0.359$ among all the models. The best performance of ATP-SDS with DF1 shown in Table~\ref{table4} achieves $\mathbf{0.390}$, outperforming the first place ($\mathbf{0.324}$) \cite{Lu2018} in the challenge by $\mathbf{6.6}$ percentage points. The embedding-level ATP-SDS with DFW achieves the best average clip-level $F_1$ score of $0.639$ among all the models.

We illustrates the average results with the confidence interval of the 20 experiments in Figure~\ref{fig7}. Since different window sizes of median filters in post-processing have a great impact on results, we show frame-level performances of all the models when window sizes are fixed with 27 and adaptive window sizes employ different $\beta$ ($\beta=\frac{1}{2}$, $\beta=\frac{1}{3}$, $\beta=\frac{1}{4}$ and $\beta=\frac{1}{5}$) respectively. As shown in in Figure~\ref{fig7}, the value of $\beta$ shows a significant effect on the frame-level performance of the model and the model tends to perform the best with $\beta=\frac{1}{3}$.

\begin{figure*}[t]
  \centering
  \subfigure[Event detection performance]{\label{fig11a}\includegraphics[width=0.96\textwidth]{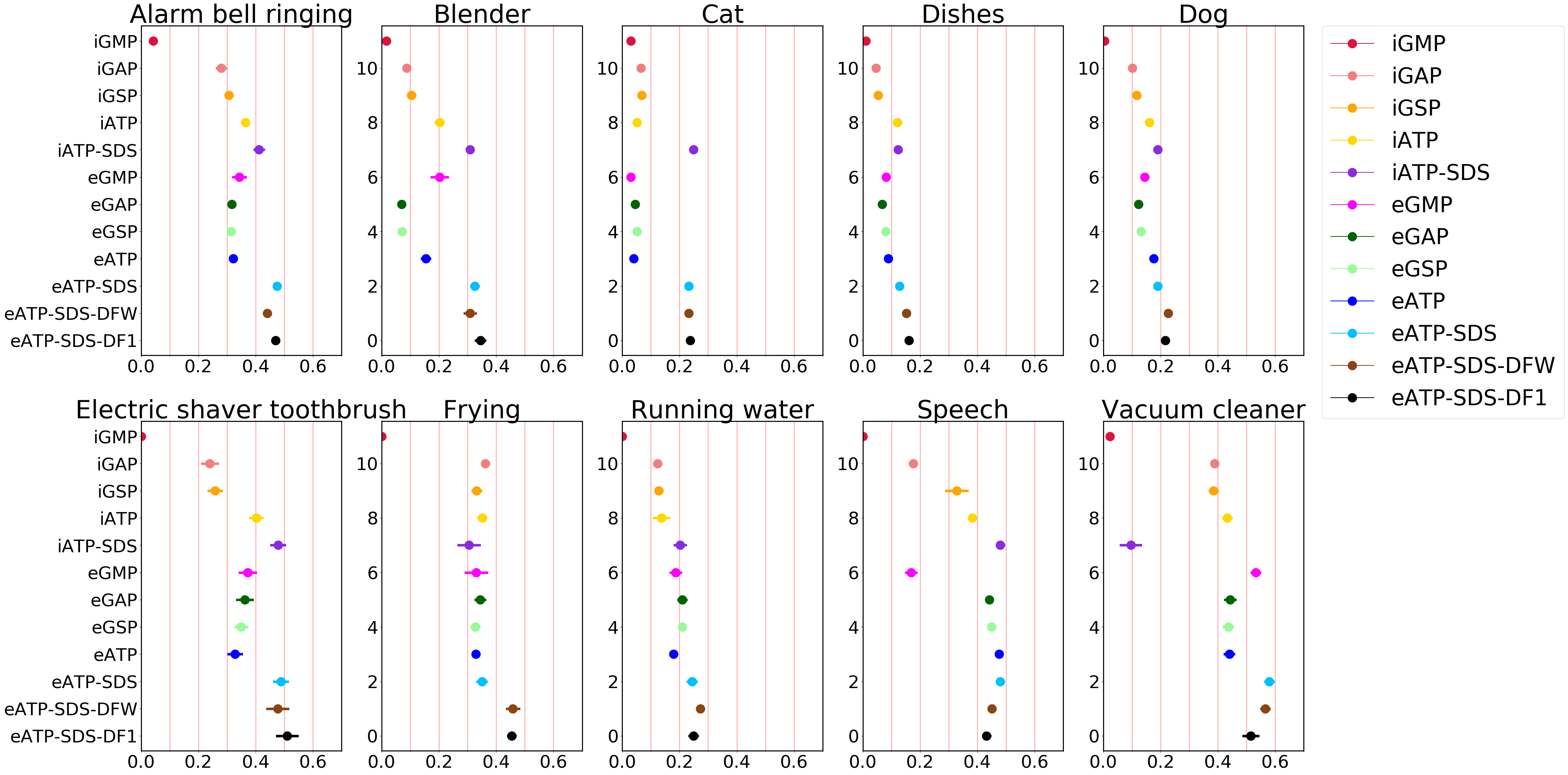}}
  \vskip 0.1in
  \subfigure[Audio tagging performance]{\label{figp11b}\includegraphics[width=0.96\textwidth]{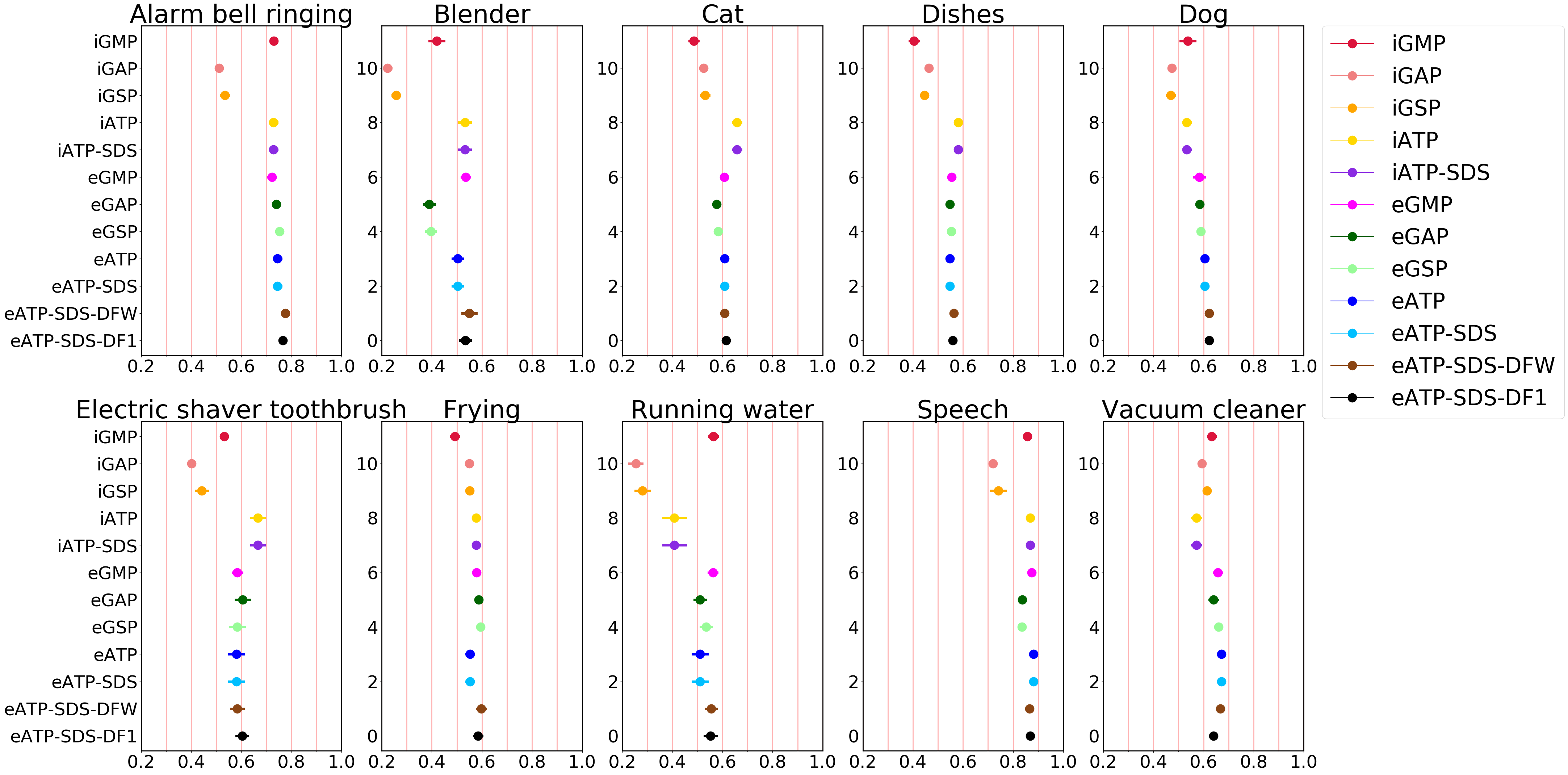}}
  \caption{The frame-level class-wise performance of models.}
  \label{fig11}
\end{figure*}

\subsection{The performances of different pooling modules}
By comparing the performances of the embedding-level models with those of the instance-level models in Table~\ref{table3}, we find that the embedding-level models outperform the instance-level models, except that the event detection performance of the embedding-level ATP is a little worse than that of the instance-level ATP. The poor event detection performance might be due to the poor ability of the shared decision surface for the embedding-level ATP. Without taking account of SDS and DF, ATP is dominant in event detection. We note that the embedding-level GMP performs best on audio tagging but worst on event detection. The best clip-level performance attributes to the fact that the strategy which GMP takes to select frames to update tends to make fewer mistakes. The relative poor frame-level performance is due to the fact that such a strategy which updates limited frames once a time leads to weak predictions of those frames that are not critical to the clip-level decision making.

As shown in Figure~\ref{fig9}, we give an example audio clip of the test set to compare the frame-level performances of the 4 embedding-level models. Dark shadows in the figure represent frame-level probabilities output by the model without smoothing, the values of which range from $0$ to $1$. Compared with the groundtruth, GMP ignores those frames that are not critical to the clip-level decision making and make extremely discontinuous predictions both for ``Blender'' and ``Speech", while the strategies that GAP and GSP take to select frames pay more attention to negative frames, leading to an incorrect prediction for ``Blender" but achieving better boundary detection for ``Speech". ATP achieves a better tradeoff between the two conditions above.

\subsection{The effect of SDS on boundary detection}
When we take SDS as decision surface for event detection, the frame-level average performance of the instance-level ATP is improved by $2.4$ percentage points and that of the embedding-level ATP is improved by $9.6$ percentage points as shown in Table~\ref{table3}. As shown in Figure~\ref{fig9}, we give an example audio clip of the test set to compare the frame-level performances of ATP and ATP-SDS. The predictions of ATP-SDS are obviously closer to groundtruths.

As shown in Figure~\ref{fig10}, we transform the high-level feature representations generated from the feature encoders of the embedding-level models into a two-dimensional space using principal components analysis (PCA) for observation. To highlight the frame-level decision surface, we only draw all the frames in positive clips (clips predicted to be positve). The yellow points represent frames predicted to be positive and purple points represent frames predicted to be negative. We can intuitively find that SDS clearly matches the boundary of two clusters  discussed in Section~\ref{Specialized decision surface}. However, the shared decision surfaces of ATP, GSP, GAP and GMP show a poor ability to separate these two clusters and lead to poor performance on event detection. Among them, the constraint that the shared decision surface of GSP is exactly consistent with its SDS hinders the flexible formation of SDS and exerts a negative effect on the separation of the two clusters. These observation exactly meets what we expect in Section~\ref{Specialized decision surface}.

It is worth mentioning that since the clear boundaries shown in Figure~\ref{fig10} separate the frames that predicted to be positive or negative rather than the groundtruths, the event detection performance is not as manifest as that of the separation of the two clusters.

%\begin{figure}[t]
%  \centering
%  \includegraphics[width=\columnwidth]{DCASE2018_class_atf1.png}
%  \caption{ The class-wise performances of $F_1$ on audio tagging of different models per category.}
%  \label{fig10}
%\end{figure}
%\begin{figure}[t]
 % \centering
%  \includegraphics[width=\columnwidth]{DCASE2018_class_f1.png}
 % \caption{ The class-wise performances of $F_1$ on event detection of different models per category.}
%  \label{fig11}
%\end{figure}

\subsection{The effect of DF on multi-label classification}
When we combine embedding-level ATP-SDS with DF, the frame-level average performance of ATP-SDS with DFW outperforms ATP-SDS by $0.9$ percentage points and that of ATP-SDS with DF1 outperforms ATP-SDS by $1$ percentage points as shown in Table~\ref{table3}. We report average frame-level $F_1$ and  average clip-level $F_1$ of 20 experiments of all $10$ event categories in Figure~\ref{fig11}.

%We argue that the high rate of co-occurrence between these categories and small scale of samples imply less identifiable information of each class, which increases the difficulty to learn better contextual represents in the high-level feature space.

%However, when we prepare a specific subspace for each category using DF, the volume of these subspaces actually are greatly reduced, making it easier to fit a small amount of data containing identifiable information. As shown in Figure~\ref{fig12}, these subspaces can be easily distinguished from each other without pre-training, which strengthens the anti-interference ability between categories of the model. As shown in \ref{fig13}, the clusters of the contextual representations with DF are more compact than those without, especially for "Frying".

%\begin{figure}[t]
%  \centering
%  \vskip 0.1in
%  \includegraphics[width=\columnwidth]{PCA_sep.png}
 % \caption{The contextual representations predicted to be positive of different categories of the test set (PCA).}
%  \label{fig12}
%\end{figure}

As shown in Figure~\ref{fig4}, ``Dishes" and ``Frying" often occur in co-occurrence with each other and have a relatively small proportion in the training set while ``Speech" often occurs in co-occurrence with any other categories. ``Running water" also often occurs in co-occurrence with ``Dishes". As shown in Figure~\ref{fig11}, ATP-SDS with DFW and DF1 did improve class-wise performances of ``Dishes", ``Frying" and ``Running water" both on event detection and audio tagging. This is because the separate subspaces of each category reduce the interference between them. However, the class-wise performance of ATP-SDS with DFW and DF1 for ``Speech" is a little worse. We argue that since the number of clips containing ``Speech" is much larger than that of clips containing any other event categories in the training set, the feature encoder without employing DF tends to form a high-level feature space more suitable for the event category ``Speech". When DF is employed, the other categories are much less disturbed by ``Speech" and make a more balanced contribution to the feature encoder, which raises overall class-wise performances but also leads a little poorer performance of category ``Speech".

Another point is that DF reduces redundant weights of the model and improve performance while improving the training efficiency. Therefore, we argue that DF is a regularization method to deal with multi-label problems using the unbalanced dataset, which is worth pushing further.

\section{Conclusions}
\label{Conclusions}
In this paper, we present how to generate frame-level probabilities for the embedding-level MIL approach and propose a specialized decision surface (SDS) and a disentangled feature (DF) for weakly-supervised polyphonic SED.

Firstly, we approach it as an MIL problem and then introduce an MIL framework with neural networks and pooling module. This framework is common in some weakly-supervised tasks and is grouped into two approaches: the instance-level approach and the embedding-level approach. We enable the embedding-level approach to make instance-level predictions and demonstrate the embedding-level approach tends to outperform the instance-level approach on experimental dataset. Based on the exploration of the ability of the embedding-level approach to produce frame-level probabilities, we propose a specialized decision surface (SDS) to detect more accurate boundaries of events.

%\begin{figure}[t]
 % \centering
%  \includegraphics[width=\columnwidth]{PCA_11_70.png}
%  \caption{The clusters of contextual representations of two different categories (PCA). Orange points represent positive clips while blue points represent negative clips per category.}
 % \label{fig13}
%\end{figure}

Secondly, to tackle the common problem causing by category co-occurrence between categories and data unbalance in the multi-label task, we propose a disentangled feature, which determines several certain subspaces for different categories without pre-training according to the prior information. In terms of optimizing the structure of the neural network, DF reduces redundant weights in the network and improves the training efficiency. From the perspective of feature space, DF optimizes the feature encoder and reduces the volume of high-level feature space of categories with insufficient samples, thus making it easier to learn more compact distribution. At the same time, DF, combined with prior information about co-occurrence between categories, reduces the interference between categories and improves the performance of the model.

Finally, the results of experiments on the dataset of DCASE2018 task 4 confirm our conjecture, which achieves a frame-level $F_1$ of $39.0\%$, outperforming the first place in the challenge by $6.6$ percentage points.

% or


\section*{Acknowledgment}
This work is partly supported by Beijing Natural Science Foundation (4172058).

\ifCLASSOPTIONcaptionsoff
  \newpage
\fi



%\bibliographystyle{IEEEtran}

%\begin{thebibliography}{1}
\bibliographystyle{IEEEtran}
\bibliography{IEEEexample}
%\end{thebibliography}


\begin{IEEEbiography}[{\includegraphics[width=1in,height=1.25in,clip,keepaspectratio]{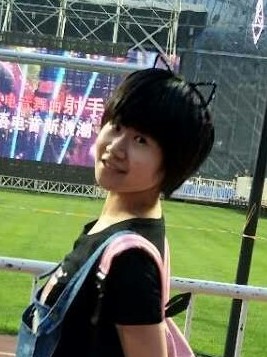}}]{Liwei Lin}
received the B.E. degree in Computer Science and Technology from China Agricultural University, Beijing, China, in 2017.
She is currently pursuing an M.E. degree in Computer Science and Technology in Institute of Computing Technology, Chinese Academy of Sciences, Beijing, China. her research interest includes audio signal processing and machine learning.
\end{IEEEbiography}

\begin{IEEEbiography}[{\includegraphics[width=1in,height=1.25in,clip,keepaspectratio]{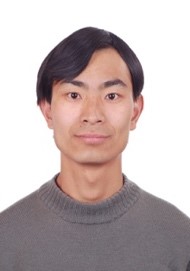}}]{Xiangdong Wang}
is an associate professor in Institute of Computing Technology, Chinese Academy of Sciences, Beijing, China. He received Doctor’s degree in Computer Science at Institute of Computing Technology, Chinese Academy of Sciences, Beijing, China, in 2007. His research field includes human-computer interaction, speech recognition and audio processing.
\end{IEEEbiography}

%\newpage

\begin{IEEEbiography}[{\includegraphics[width=1in,height=1.25in,clip,keepaspectratio]{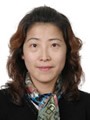}}]{Hong Liu}
is an associate professor in Institute of Computing Technology, Chinese Academy of Sciences, Beijing, China. She received her Doctor’s degree in Computer Science at Institute of Computing Technology, Chinese Academy of Sciences, Beijing, China, in 2007. Her research field includes human-computer interaction, multimedia technology, and video processing.
\end{IEEEbiography}

\begin{IEEEbiography}[{\includegraphics[width=1in,height=1.25in,clip,keepaspectratio]{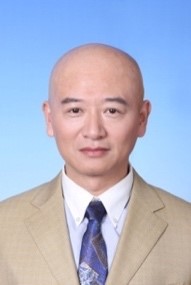}}]{Yueliang Qian}
is a professor in Institute of Computing Technology, Chinese Academy of Sciences, Beijing, China.
He received his Bachelor’s degree in Computer Science at Fudan University, Shanghai, China in 1983. His research field includes human-computer interaction and pervasive computing.
\end{IEEEbiography}

%\vfill


\end{document}